\documentclass[aps,notitlepage,preprintnumbers,onecolumn,amsmath,amssymb,floatfix,superscriptaddress,pra]{revtex4-1}

\pdfoutput=1 
\usepackage[utf8]{inputenc}

\usepackage{amsmath}
\usepackage{amsfonts}
\usepackage{braket}
\usepackage{graphicx}
\usepackage[usenames,dvipsnames,svgnames,table]{xcolor}

\usepackage[colorlinks=true,linkcolor=blue, citecolor=blue, bookmarks]{hyperref}

\def\Tr{\mbox{Tr}\,}

\def\be{\begin{equation}}
\def\ee{\end{equation}}

\def\bea{\begin{eqnarray}}
\def\eea{\end{eqnarray}}

\def\e{\epsilon}

\def\Tr{\mbox{Tr}\,}

\begin{document}
\title{Entanglement revivals  as a probe of scrambling in finite quantum systems}
\author{Ranjan Modak} 
\affiliation{SISSA and INFN, via Bonomea 265, 34136 Trieste, Italy}
\author{Vincenzo Alba}
\affiliation{Institute  for  Theoretical  Physics,  Universiteit  van  Amsterdam,Science  Park  904,  Postbus  94485,  1098  XH  Amsterdam,  The  Netherlands}
\author{Pasquale Calabrese}
\affiliation{SISSA and INFN, via Bonomea 265, 34136 Trieste, Italy}
\affiliation{International Centre for Theoretical Physics (ICTP), Strada Costiera 11, 34151 Trieste, Italy}

\begin{abstract}
The entanglement evolution after a quantum quench  became one of the tools to distinguish integrable versus chaotic (non-integrable) quantum many-body dynamics.
Following this line of thoughts, here we propose that the revivals in the entanglement entropy provide a finite-size diagnostic benchmark for the purpose. 
Indeed, integrable models display periodic revivals manifested in a dip in the block entanglement entropy in a finite system. 
On the other hand, in chaotic systems, initial correlations get dispersed in the global degrees of freedom (information scrambling) and such a dip is suppressed. 
We show that while for integrable systems the height of the dip of the entanglement of an interval of fixed length decays as a power law with the 
total system size, upon breaking integrability a much faster decay is observed, signalling strong scrambling. 
Our results are checked by exact numerical techniques in free-fermion and free-boson theories, and 
by time-dependent density matrix renormalisation group in interacting  integrable and chaotic models. 

\end{abstract}
\maketitle

\section{Introduction}
\label{sec:intro}

Integrable and chaotic many-body quantum systems are very different objects. 
The former have infinite classes of local charges constraining their dynamics;
conversely, the latter have only few local integrals of motion. 
Integrable systems admit stable quasiparticles with infinite lifetime and elastic scattering between them \cite{m-book}.
When non-integrable models have some quasiparticles at all, they have a finite lifetime and inelastic scattering with particles production. 
One would then expect their unitary non-equilibrium dynamics (say following a quantum quench \cite{cc-06}) to be completely different \cite{calabrese-2016}. 
Indeed, it is nowadays well established that while chaotic systems long time after a quench 
are described by the Gibbs (thermal) ensemble~\cite{v-29,eth0,deutsch-1991,srednicki-1994,rdo-08,rigol-2012,dalessio-2015,gogolin-2015},
an integrable system fails to thermalise and a statistical description of its asymptotic local properties 
requires a Generalised Gibbs Ensemble (GGE)~\cite{rigol-2007,cazalilla-2006,barthel-2008,cramer-2008,cramer-2010,scc-09,cazalilla-2012a,calabrese-2011,cc-07,mc-12,sotiriadis-2012,collura-2013,fagotti-2013,p-13,fe-13b,fagotti-2013b,sotiriadis-2014,fcec-14,ilievski-2015a,ilievski-2016b,alba-2015,langen-15,essler-2015,cardy-2015,sotiriadis-2016,bastianello-2016,vernier-2016,pvw-17,vidmar-2016,ef-16,fgkc-17,pk-17,ilievski-2016}, which takes into account all the local and quasilocal conserved quantities of the system~\cite{ilievski-2015a,ilievski-2016b,ilievski-2016}.
However, there are many cases when the predictions from GGE and thermal ensembles are too close to each other. 
Furthermore, an integrable point constrains the dynamics of a relatively large neighbourhood   in parameter space, giving rise to the phenomenon of 
prethermalisation, according to which only for extremely large times the thermal behaviour  is attained, while at accessible times one observes a quasi-stationary 
state with quasi-integrable features \cite{pret,MarcuzziPRL13,EsslerPRB14,konik14,BF15,BEGR:PRL,FC15,mmgs-16,af-17}.
 
These observations were among those that initiated the search for concepts that naturally take integrable and chaotic models apart. 
A very suggestive proposal concerns the {\it scrambling of quantum information}, hence of entanglement. 
Indeed, qualitatively we expect that in integrable systems the  spreading of entanglement, due to the quasiparticles with a local-in-space structure,  
happens in a ``localised'' manner, i.e., local initial correlations are somewhat preserved.  
This is the essence of the quasiparticle picture for entanglement spreading \cite{calabrese-2005,alba-2016,alba.2018,c-18}. 
This scenario must change in non-integrable models: the common lore is that inelastic processes and the quasiparticle 
decay should facilitate the delocalisation of the initial correlation in the global degrees of freedom of the 
system \cite{zanardi-2001,prosen-2007,prosen-2007a,znidaric-2008,pizorn-2009,dubail-2017,maldacena-2016,bll-18,
khemani-2018,sarang-2018}. This idea originated in the context of the black-hole 
information paradox ~\cite{hayden-2007,sekino-2008,shenker-2014}. There the initial 
correlation between the interior and the exterior of the black hole gets dispersed in the global degrees of freedom of 
the radiation that remains after the black hole evaporates. 

Unfortunately, probing the scrambling scenario in microscopic quantum many-body systems proved to be a daunting task. 
Several diagnostic tools have been devised, such as the tripartite mutual information~\cite{hosur-2016,pappa-18}, 
out-of-time-ordered correlators~\cite{shenker-2014}, and operator space 
entanglement entanglement entropy~\cite{adm-2019,bkp-20,nnrt-19}.
An idea that is related to this paper is to diagnose scrambling from the time evolution of the mutual information between two disjoint intervals. 
Indeed, in integrable models, because of the infinite lifetime of the quasiparticles, 
the mutual information exhibits a peak at intermediate times, as first pointed out in conformal field theory (CFT) \cite{calabrese-2005}.  
For chaotic systems, due to scattering and finite lifetime of quasiparticles, the peak should decay, signalling  scrambling. 
This behaviour  has been first proposed and observed in CFTs with large central charge~\cite{asplund-2014,bala-2011,asplund-2015,leich-2015}
and better characterised   for integrable systems with non-linear dispersion in Ref.~\cite{alba.2019}.

It is of fundamental importance to device further diagnostic tools to probe quantum information scrambling. 
Here, we propose and show that entanglement revivals in finite systems can be used for this purpose. 
The idea to use revival effects to distinguish between ergodic  and non-ergodic systems has already been explored in Refs.~\cite{moore.2015,mtpas-20}, but in different contexts.
Revivals have been studied in Conformal Field Theory~\cite{cardy-2014,stephan-2011} 
(see Ref.~\onlinecite{najafi-2017,najafi-2019} for some comparison with lattice realisations). 
Nevertheless, the inspiration for our work came from recent results for maximally chaotic models in which the entanglement entropy shows
no revivals at all in finite size \cite{bkp-19,pbcp-20,cdc-18}, while in the integrable limit a perfect recurrence is observed. 
These results are the two black and white extremes of a more refined and grey structure which we investigate and characterise  here. 
 
Our main result is that  it is possible to detect scrambling by monitoring the first entanglement revival. 
We consider a subsystem of fixed size $\ell$ embedded in a system of finite size $L$. 
We consider initial product states, i.e., with zero entanglement entropy. 
For time $t\lesssim\ell$ the entropy increases linearly. At larger times $\ell\ll t\lesssim L$ 
the entropy saturates. At the first revival time $t=t_\mathrm{R}\propto L$, 
the entanglement entropy decreases, and it exhibits a dip, which is followed by a later increase. 
As $L$ increases the revival time $t_\mathrm{R}\propto L$ is shifted to longer times and the 
height of the dip decreases. 
We show that for integrable systems the dip decays algebraically in $L$ (as $L^{-1/2}$ within the quasiparticle picture), as 
we explicitly test in free-fermion, free-boson, and interacting integrable models. 
Remarkably, upon breaking integrability, the dip exhibits a much faster decay, signalling strong scrambling.

A main observation concerns the comparison with the mutual information, 
which is a very similar diagnostic tool for scrambling that two of us 
proposed in Ref.~\onlinecite{alba.2019}. For the entanglement revival 
we require only to calculate the single interval entanglement entropy, 
which in the framework of Matrix Product States (MPS) is much easier 
and less numerically demanding than the entanglement entropy of 
two intervals required for the mutual information. As a drawback, 
we need to explore times growing linearly with $L$ and, depending 
on circumstances, such times can be harder than those required for 
the mutual information that must grow linearly with the separation 
of the two intervals. 
At the end of the day, we believe that the combined use of the two 
tools provides a rather clear indication of the integrability/chaotic nature
of the model under scrutiny.

The paper is organised  as follows. 
In Sec.~\ref{sec-I} we introduce the scrambling of entanglement revivals in integrable models within the quasiparticle picture. 
In Secs.~\ref{sec II} and \ref{sec IIc} we work out explicit predictions and test them for free-fermion and free-boson models, respectively. 
In Sec. \ref{sec IId} we move to interacting integrable models focusing on the Heisenberg spin chain
and a numerical study of its quench dynamics by means of time-dependent Density Matrix Renormalisation  Group (tDMRG) techniques. 
In section~\ref{sec IIIc} we discuss the effects integrability breaking on the entanglement revivals. 
Our conclusions are in section~\ref{sec IV}.



\section{Scrambling and entanglement revivals in integrable models}
\label{sec-I}

The typical  protocol to drive a system out of equilibrium is the {\it quantum quench}~\cite{cc-06}: 
an isolated system is initially prepared in a non-equilibrium pure state $|\psi_0\rangle$ 
and then it is let evolve under the unitary dynamics governed by a Hamiltonian $H$. 
Although the dynamics is unitary and the entire system remains in a pure state, 
at long times the reduced density matrix $\rho _A$ of an arbitrary finite 
compact subsystem $A$ of length $\ell$ displays local thermodynamic equilibrium. 
The reduced density matrix  $\rho_A$ is defined as 
\begin{equation}
\label{rhoA}
\rho_A\equiv\textrm{Tr}_B|\psi(t)\rangle\langle\psi(t)|, 
\end{equation}
where the trace is over the degrees of freedom of the complement  $B$ of $A$, 
and $|\psi(t)\rangle\equiv e^{-i H t} |\psi_0\rangle$ is the time-dependent state of the system.
Local thermal equilibrium means that $\rho_A$ for large times equals the reduced density matrix of a statistical ensemble \cite{ef-16}.
 
A key question is how entanglement spreads in out-of-equilibrium many-body 
system, both theoretically and experimentally, thanks to recent progress on the direct measurement of entanglement  dynamics
with cold atoms and trapped ions \cite{islam-2015,kaufman-2016,daley2012,elben2018,vermersch2018,brydges2019,exp-lukin}.
One of the most useful entanglement measures is the von Neumann (entanglement) entropy \cite{rev-enta}
\begin{equation}
S_{\ell}\equiv-\textrm{Tr}\rho_A\ln\rho_A.
\end{equation}
Here we are interested in the out-of-equilibrium dynamics of $S_\ell$ after a quantum quench. Our main result is that 
the dynamics of $S_\ell$ in finite-size systems depends dramatically on whether the Hamiltonian is integrable  or chaotic. 

To understand why this is the case, let us first briefly describe the 
quasiparticle picture for the entanglement spreading which is applicable to generic integrable models (first introduced in Ref.~\cite{calabrese-2005} in the context of CFT).
According to this picture, the initial state acts as a source  of quasiparticle excitations which are produced in {\it pairs} and uniformly in space. 
After being created, the quasiparticles move ballistically through the system with opposite velocities. 
Only quasiparticles created at the same point in space are entangled and while they move far apart they carry entanglement and correlation in the system. 
A pair contributes to the entanglement entropy at time $t$ only if one particle of the pair is in $A$ (the interval of length $\ell$) and its partner in $B$.
Keeping track of the linear trajectories of the particles, it is easy to conclude~\cite{calabrese-2005,alba-2016} 
\begin{equation}
\label{semi-cl}
S(t)= \sum_n\Big[ 2t\!\!\!\!\!\!\int\limits_{\!2|v_n|t<\ell}\!\!\!\!\!\!
dk v_n(k)s_n(k)+\ell\!\!\!\!\!\!\int\limits_{2|v_n|t>\ell}\!\!\!\!\!\!
dk s_n(k)\Big].
\end{equation}
Here the sum is over the species of particles $n$ whose number depends on the model, $k$ represents their quasimomentum (rapidity), 
$v_n(k)$ is their velocity, and $s_n(k)$ their contribution to the entanglement entropy. 
(Often we will work with a single species of quasiparticle omitting the sum over $n$ and the subscripts). 
The quasiparticle prediction~\eqref{semi-cl}  for the entanglement entropy holds true in the space-time 
scaling limit, i.e. $t,\ell\to \infty$ with the ratio $t/\ell$ fixed.  
When a maximum quasiparticle velocity $v_M$ exists (e.g., as a consequence of the Lieb-Robinson bound~\cite{lieb-1972}), Eq.~\eqref{semi-cl} predicts 
that for $t\le \ell/(2v_M)$, $S_{\ell}$ grows linearly in time. 
Conversely, for $t\gg\ell/(2v_M)$, only the second term survives and the entanglement is extensive in the subsystem size, i.e., $S_{\ell}\propto\ell$. 
In order to give predictive power to Eq. \eqref{semi-cl}, one should fix the values of $v_n(k)$ and $s_n(k)$:
the former are the group velocities of the excitations around the stationary state~\cite{BEL-14,alba-2016,alba.2018} 
and the latter are the thermodynamic entropy densities of the GGE ~\cite{alba-2016,alba.2018}.
The validity of Eq.~\eqref{semi-cl} has been carefully tested both analytically and numerically in free-fermion and free-boson  
models~\cite{calabrese-2005,fagotti-2008,ep-08,nr-14,kormos-2014,leda-2014,collura-2014,bhy-17,hbmr-17,buyskikh-2016,fnr-17,mm-20,knn-19,dat-19} and in many interacting 
integrable models~\cite{alba-2016,alba.2018,PVCP18_I, PVCP18_II,MBPC17,modak.2019,dmcf-06}. 
The mechanism for the entanglement evolution in chaotic systems is different, not as well understood as in integrable models and with many peculiar features. 
Nevertheless, the entanglement entropy grows linearly up to a time extensive in subsystem 
size \cite{bkp-19,pbcp-20,cdc-18,hk-13,nahum-17,lauchli-2008,nahum-18,nahum-18b,r-2017,kctc-17,mkt-18}, exactly as in integrable systems.

Interestingly, Eq.~\eqref{semi-cl} has been generalised to capture the non-equilibrium dynamics in many different physical situations and to other physical quantities. 
For example we mention quenches from inhomogeneous initial states~\cite{,alba-inh,bertini-inh,alba-inh.2019,alvise.2019,dsvc17,ma-20,ctd-19,rbd-19} and
states with more complicated quasiparticle structure than simple uncorrelated pairs \cite{btc-18,bc-18,bc-20}.
Furthermore, the picture has been adapted to describe the steady-state R\'enyi entropies~\cite{alba_renyi_qa.2017,alba_renyi.2017,alba_renyi.2019,mestyan.2018}  
and to the dynamics of the logarithmic negativity~\cite{coser-2014,alba_qi.2018,knrt-19}. 
Very recently, it has been shown that by using the quasiparticle picture it is also 
possible to study the fate of the entanglement in free-fermion systems in the presence of dissipation~\cite{alba-2020} (see also Ref.~\onlinecite{somnath-2020}). 
An aspect that is of much importance for our analysis is the behaviour of the mutual information between two disjoint intervals \cite{alba.2019}.
In integrable systems, the mutual information exhibits an algebraic decay with the distance between the intervals, that  
within the quasiparticle picture has a decay exponent equal to $1/2$.
Away from the scaling limit, the power-law behaviour persists, but with a larger (and model-dependent) exponent. 
For non-integrable models, a much faster decay, compatible with an exponential,  is observed.

\begin{figure}
\includegraphics[width=0.55\textwidth] {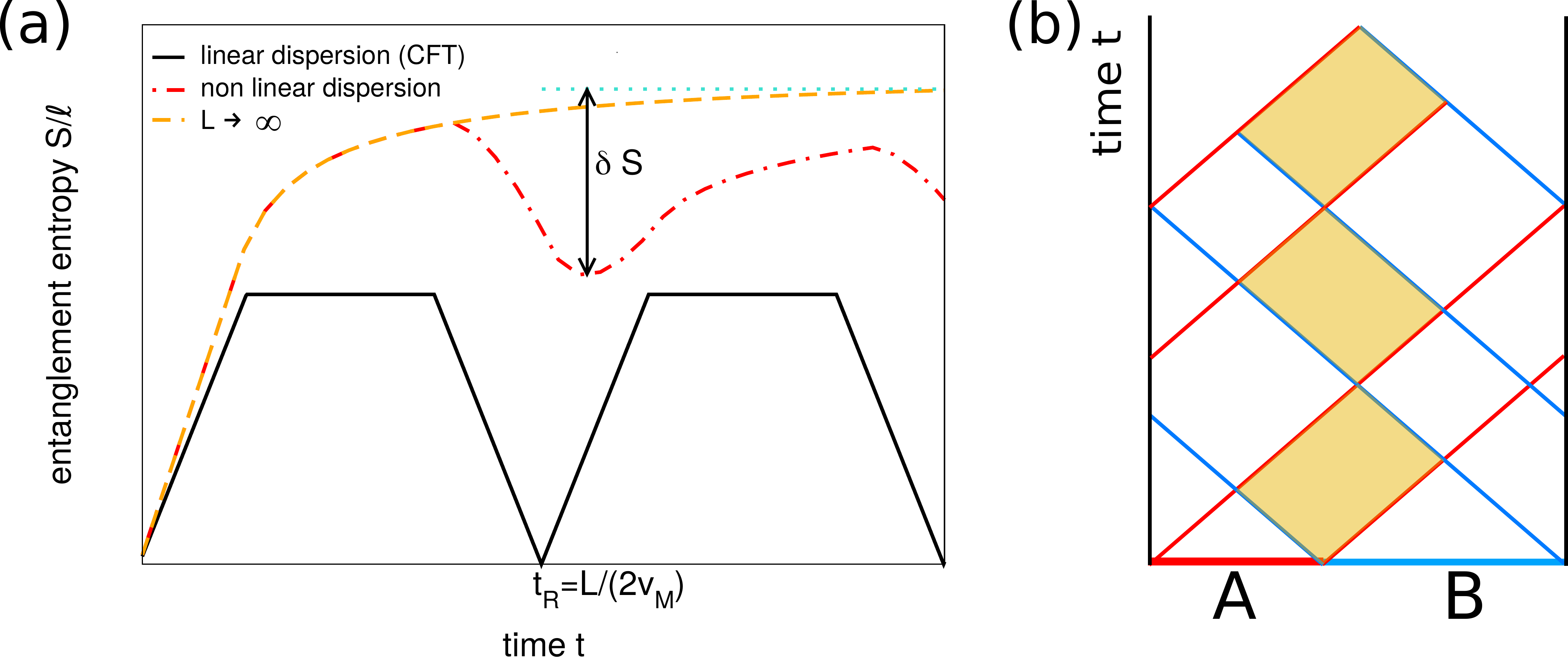}
\caption{(a) Schematic diagram of entanglement dynamics for finite-size integrable systems.
The continuous line is the quasiparticle prediction for a model with exactly linear dispersion, such as a CFT. 
Note the perfect revival at $t_R$. 
The dashed-dotted line is the quasiparticle prediction for a model with a realistic nonlinear dispersion. 
For $L\to\infty$ the entropy saturates (dashed line), while at finite size (dot-dashed line) it shows a dip of height $\delta S$, which is the main quantity we consider. 
The precise definition of $\delta S$ is shown pictorially as the difference between the asymptotic result for infinite systems ($S_\ell(\infty)$, dot line)
and the minimum of the entropy close to the first revival. (b) Pictorial interpretation 
of~\eqref{ee}. The slanted lines are the trajectories of the entangled quasiparticles created at the interface between the two subsystems. 
At a given time $t$, the entanglement entropy is proportional to the horizontal section of the shaded region at that time.
Three revivals are explicitly shown. The quasiparticles turn around the system because of periodic boundary conditions
}
\label{fig0}
\end{figure}

We recall that in Eq.~\eqref{semi-cl} we assume that subsystem $A$ of length $\ell$ is embedded in an infinite system. 
The result is different for a finite system of total length $L$ in which, as we are going to show, the entanglement entropy exhibits revivals (we focus on periodic boundary conditions, 
but other boundary conditions lead just to minor adjustments as long as they are compatible with integrability; indeed we will later turn to open boundary conditions too).
We assume that $L$ is large enough so that quasiparticles are well-defined;
this implies that the space-time scaling limit generalises as $t,\ell,L\to\infty$ with both $t/\ell$ and $\ell/L$ fixed (or, equivalently, $t/L$).
At this point, to study  the revivals we need to modify Eq.~\eqref{semi-cl}, simply accounting for the quasiparticles trajectories in a periodic system.

\begin{figure}
\includegraphics[width=0.55\textwidth] {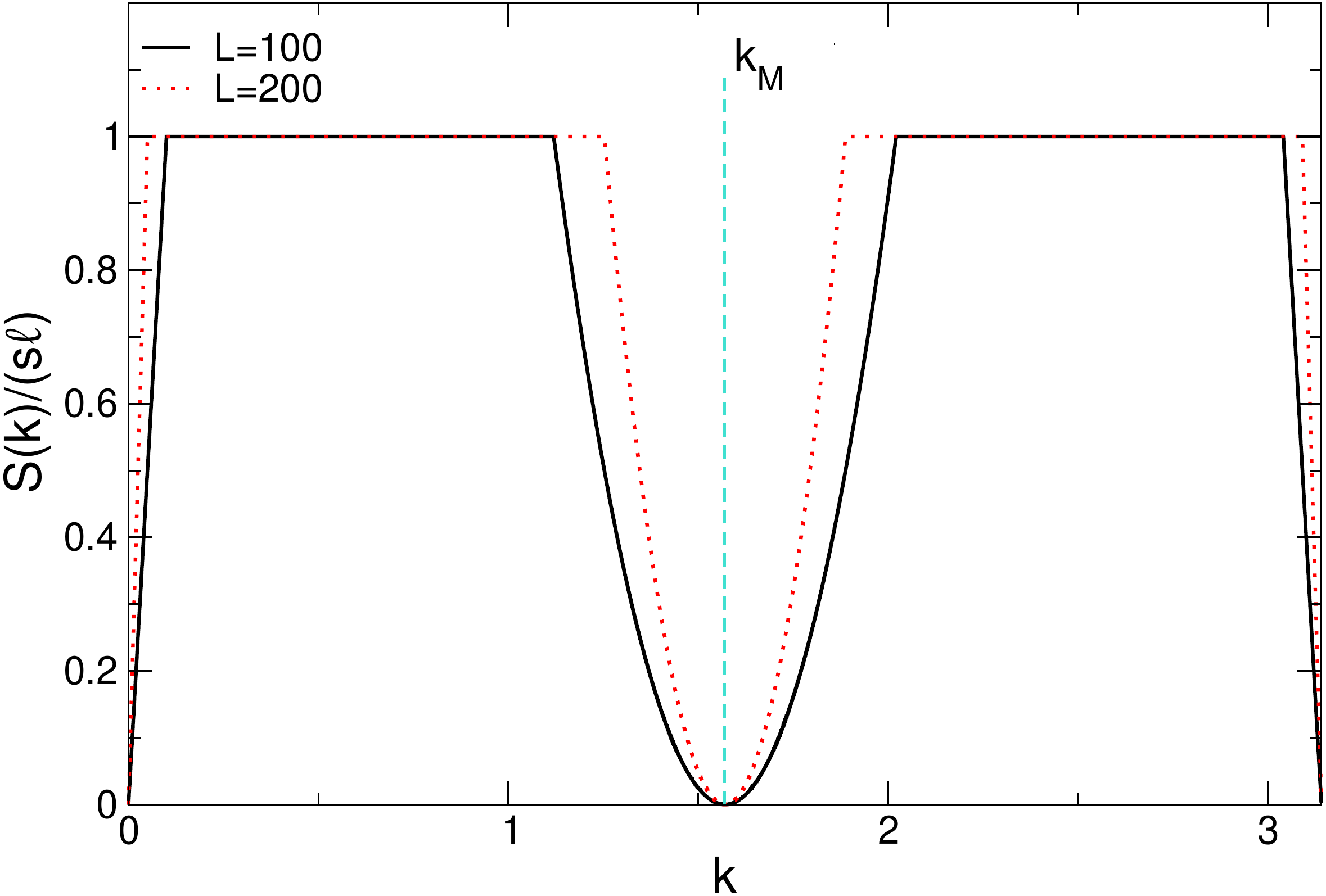}
\caption{ Physical mechanism for the vanishing of the entanglement revivals in integrable 
 systems. The figure shows the contribution of the quasiparticles 
 to the density of entanglement entropy $S(k)/(s\ell)$ at the revival time $t=L/(2v_{M})$ 
 plotted versus their quasimomentum $k$. Here we consider quasiparticles with 
 group velocity $v(k)=\sin(k)$, assuming $S(k)/\ell=s$, with $s$ a constant. 
 The entropy density is obtained by 
 integrating over $k$. In the limit $L\to\infty$ with finite 
 subsystem size $\ell$ all the quasiparticles contributions are present, i.e., 
 $S(k)/(s\ell)=1$ for any $k$. At finite $L$ the contributions due to slow quasiparticles 
 (with $k\approx 0$ and $k\approx\pi$) and quasiparticles with maximum velocity at 
 $k\approx k_M$ are suppressed (dips in the figure). This gives rise to the dip 
 in Fig.~\ref{fig0} (a). Note that in the limit $L\to\infty$ the size of the 
 dips due to the missing quasiparticles shrinks, reflecting the vanishing of the 
 entanglement revival. Note also that the entanglement revival 
 is dominated by quasiparticles with  maximum velocity. 
}
\label{fig0a}
\end{figure}

Let us start by describing what happens when all quasiparticles have the same velocity $v$ regardless of momentum, as it is the case, e.g., in CFT.
The entropy first grows linearly up to $\ell/(2v)$. Then, it exhibits a plateau. Up to this time, the entanglement behaves like in an infinite system. 
The plateau terminates at $t= (L-\ell)/(2v)$ when the first quasiparticles produced at one boundary of the 
subsystem re-enter into it from the other edge after having turned around the circle.
After this time, more and more quasiparticles will re-enter the subsystem producing a drop of the entropy that lasts until the {\it revival  time} $t_R=L/(2v)$, 
when the dynamics restart exactly like if the system was at $t=0$. 
And so on in a periodic fashion in time. 
This entropy dynamics is shown in  Fig.~\ref{fig0} (a) as a full line.  
The physical origin of the entanglement revivals is illustrated in Fig.~\ref{fig0} (b), focusing 
on the trajectories of the quasiparticles. To understand the entanglement dynamics is sufficient to consider 
the trajectories of the entangled quasiparticles created at the interfaces between $A$ and $B$ (slanted lines 
in Fig.~\ref{fig0} (b)). Note that in the figure we use periodic boundary conditions. It is clear 
that at a given time $t$ the number of entangled pairs that are shared between $A$ 
and $B$, and hence the entanglement entropy, is proportional to the horizontal section at that time 
of the shaded area in the figure.

Taking into account the different velocities $v(k)$ of the quasiparticles is now trivial:  
it is enough to sum/integrate over all possible quasi-momenta $k$ the result for each mode.
For the case of a single species of quasiparticles, one obtains 
\begin{multline}
S_{\ell}(t)=\int_{\left\{\frac{2v(k)  t}{ L}\right\} < \frac{\ell}{L} }{\frac{dk}{2\pi} s(k) L\left\{\frac{2v(k)  t}{ L}\right\}} +\ell \int_{ \frac{\ell}{L}\leq \left\{\frac{2v(k)  t}{ L}\right\} < 1-\frac{\ell}{L}}{\frac{dk}{2\pi} s(k)}  \\
+\int_{ 1-\frac{\ell}{L}\leq\left\{\frac{2v(k)  t}{ L}\right\} }{\frac{dk}{2\pi} s(k) L\left(1-\left\{\frac{2v(k)  t}{ L}\right\}\right)} ,
\label{ee}
\end{multline}
where $\{x\}$ denotes the fractional part of $x$, e.g.,  $\{ 1.76\}=0.76$.  
This result has been already reported in Ref.~\cite{lucas.2019} (suggested by one of the present authors, as acknowledged there), but in a different context. 
The structure of  Eq.~\eqref{ee} is the same  as Eq.~\eqref{semi-cl}: $s(k)$ is the thermodynamic entropy of the GGE that describes the steady state, and 
$v(k)$ are velocities of the low-lying excitations around the corresponding thermodynamic macrostate. 
Taking into account the existence of more species trivially amounts to sum also over the different species,  each with its own velocity and entropy density
in momentum space, in full analogy to Eq.~\eqref{semi-cl}.

We now discuss the time evolution of $S_\ell$ for fixed $\ell,L$ as predicted by Eq. \eqref{ee}, for systems with nonlinear quasiparticles dispersion. 
The qualitative behaviour is shown as a dashed-dotted line in Fig.~\ref{fig0} (a). 
The entropy grows linearly up to $\ell/(2v_M)$, where now $v_M$ is the maximum quasiparticles velocity. 
Importantly, in a realistic quantum many-body system there is a continuum of excitations with different velocities. 
This implies that  the ``plateau'' in Fig.~\ref{fig0} (a) at $\ell/(2v_M)\ll t\lesssim (L-\ell)/(2v_M)$ is not flat  
but shows a slow saturation in the limit $L\to\infty$. 
Such slow saturation terminates at $t= (L-\ell)/(2v_M)$. 
After  the most important aspect for our purposes comes: at $t=t_R\equiv L/(2v_M)$ the entropy exhibits a {\it partial revival}, 
in contrast to the case with a single velocity when the revival is perfect. 
The reason for this partial revival is obvious: when the fastest quasiparticles do not entangle anymore $A$ and $B$ because the pairs are back to their 
original position, there are still many slow quasiparticles that are still contributing to $S_\ell$.
Here we focus on the dip in the entanglement entropy at (or close to) $t=t_R$. 
We consider the scaling for large $L$ at fixed $\ell$ (which is still considered large enough for the quasiparticle picture to apply),
i.e., we work in the regime $t,\ell,L\to\infty$ with $t/\ell$ and $t/L$ fixed but $\ell\ll L$.
(If we had considered the limit with $\ell/L$ of order 1, we would have approached a scaling curve for large $L$.)
See below for a discussion about the applicability of these limits. 
The dip at $t_R$ becomes less and less pronounced as $L$ gets larger (at fixed $\ell$), because more and more quasiparticles do not have time to reach the subsystems after 
going around the circle. 
To quantify this effect,  we consider the (normalised) height $\delta S$ of the dip (see Fig.~\ref{fig0} (a)), which 
is measured with respect to the saturation value of the entropy in the infinite system as
\begin{equation}
\label{eq:delta-def}
\delta S\equiv \frac{S_\ell(\infty)-S_\ell(t_R)}{\ell}, \qquad\textrm{with}\quad S_\ell(\infty)\equiv\lim_{t\to\infty} \lim_{L\to\infty}S_\ell(t). 
\end{equation}
The definition of $\delta S$ is illustrated pictorially in Fig.~\ref{fig0} (a). 

We now discuss the behaviour of $\delta S$ in the framework of the quasiparticle picture. 
We consider a generic integrable quantum many-body systems, with a continuum of entangling quasiparticles. 
Since we focus on the first revival at $t_R=L/(2v_M)$, we replace $t\to t_R$ in 
Eq.~\eqref{ee}, which yields   
\begin{equation}
S_{\ell}(t=t_R)=\int_{\frac{v(k) }{v_M} < \frac{\ell}{L} }{\frac{dk}{2\pi} s(k) L\frac{v(k) }{ v_M}} +\ell \int_{ \frac{\ell}{L}\leq \frac{v(k) }{ v_M} < 1-\frac{\ell}{L}}{\frac{dk}{2\pi} s(k)}
+\int_{ 1-\frac{\ell}{L}\leq \frac{v(k) }{ v_M} }{\frac{dk}{2\pi} s(k) L\left(1-\frac{v(k) }{ v_M}   \right)}.
\label{ee3}
\end{equation}
We now show that Eq.~\eqref{ee3} implies that $\delta S$ (cf.~\eqref{eq:delta-def}) scales like $\delta S\propto L^{-1/2}$ for $L\to\infty$. 
In fact, in the limit $L\to\infty$ (i.e. $L\gg\ell$), the dip of the entanglement entropy, as discussed above, is dominated by the quasiparticles with $v\approx v_M$
(slowest quasiparticles did not have time to go around the system).
This is illustrated in Fig.~\ref{fig0a}. In the figure we plot the 
quasiparticles contributions to the entropy density $S/\ell$ at the revival time 
$t=L/(2v_M)$ versus the momentum of the quasiparticles $k$. The result is 
obtained by using~\eqref{ee}. We restrict to quasimomenta in $[0,\pi]$ because 
the entropy is an even function of $k$. 
For simplicity we assume that $s(k)=s$ (cf.~\eqref{ee}) for any $k$ and 
$v(k)=\sin(k)$. In the limit $L\to\infty$ all the quasiparticles contribute 
to the entropy. This corresponds to the the plateaux at $t\to\infty$ in 
Fig.~\ref{fig0} (a). For finite $L$ the contributions of the slow quasiparticles 
at $k\approx 0,\pi$ and of the quasiparticles with maximum velocity $v_M=v(k_M)$ 
are suppressed. This manifest itself in the dips at $k\approx k_M$ 
and $k\approx 0,\pi$. This gives rise to the dip at the time 
of the entanglement revival in Fig.~\ref{fig0} (a). Importantly, upon increasing 
$L$ the dips in Fig.~\ref{fig0a} (a) shrink, which correspond to the vanishing of 
the entanglement revival. Note that the leading contribution in powers of 
$1/L$ originates from the fast quasiparticles with $v\approx v_M$, whereas 
the contribution of slow ones with $k\approx 0, \pi$ is subleading.

Therefore, we expand the velocity of quasiparticle $v(k)$  around $k_M$ (the momentum with maximum velocity, $v_{k_M}=v_M$) up to second order in $k-k_M$ as 
\begin{equation}
	\frac{v(k)}{v_M}=1-\frac{v''}{2v_M}(k_M-k)^2+o{(k_M-k)^3}, \quad\mathrm{with}\, 
	v''\equiv-\frac{\partial^2v(k)}{\partial k^2}\Big|_{k=k_M}. 
\end{equation}
If there are more momenta with maximum velocity, we should just sum over them. 
We assume that the entanglement content of the quasiparticles $s(k)$ is such that $s(k)=s_M+o(1)$ with a nonzero $s_M\equiv s_{k_M}$.
Plugging these expansions in Eq.~\eqref{ee3}, we obtain
\begin{equation}
	S_{\ell}(t_R)= S_\ell (\infty) - \frac{4}{3\pi}\Big(\frac{2v_M}{v''}\Big)^{1/2} 
	\Big(\frac{\ell}{L}\Big)^{1/2}s_M \ell + \ell O\Big(\frac{\ell}L\Big), 
\label{ee_main}
\end{equation}
where $S_\ell(\infty)$ is the asymptotic value of entanglement entropy in the thermodynamic limit, cf. Eq. \eqref{eq:delta-def}. 
Few comments are needed. First of all, we only considered one species of quasiparticles, but  for the validity of Eq. \eqref{ee_main} this is unimportant:
when there are more types of quasiparticles, these have well separated maximum velocities and so for large $\ell$ and $L$ only one matters 
and Eq. \eqref{ee_main} remains valid. However, the different species of quasiparticles can give strong finite size effects, as we shall see.     
Second, we note that Eq.~\eqref{ee_main} has the same structure as the formula describing the decay of the mutual information peak between two far apart 
intervals that has been derived in Ref.~\cite{alba.2019} and indeed has the same physical origin. 
Finally, let us stress the main limitation of Eq.~\eqref{ee_main}. We obtained it in the quasiparticle picture in the limit  $\ell/L\to0$.
However, this is not fully justified because the quasiparticle picture is valid for $\ell/L$ fixed and finite.   
Still, we can think of the limit in which $\ell=a L$ with $a\ll 1$. 
Thus, we expect the quasiparticle picture to describe an intermediate regime between very small $\ell$ (in which $\ell/L$ is not of order 1)
and the scaling regime with $\ell$ of the same order of $L$. 
We recall that this is exactly what it has been found for the mutual information peak in~\cite{alba.2019}.
There is not a general approach to understand the scaling for very small $\ell$ since it is expected to be model dependent;
for example, from the analogous result of the mutual information peak~\cite{alba.2019}, we would expect  for free fermion the dip to decay as $L^{-2/3}$
while for free bosons as $L^{-1}$.

We want finally to comment on another interesting aspect of Eq. \eqref{ee_main}. 
If we expanded for large $L$ and at fixed $\ell$, the quasiparticle prediction for the entanglement evolution~\eqref{ee3} at any time away from the revivals, 
we would trivially obtain an analytic behaviour in $L^{-1}$, because the velocity, away from the maximum has a linear term in $k$.
In this respect, the non-analytic scaling at the revival time is a straightforward consequence of expanding close to an extreme of the velocity
(exactly like in Landau-Ginzburg approach, the mean field exponent $1/2$ is caused by expanding close to the minimum of the free energy).

In the following section, we numerically verify Eq.~\eqref{ee_main} in free-fermion and free-boson models, 
as well as in interacting integrable models. We will show how for small $\ell$ there is a crossover towards another power-law scaling. 
Finally, we will provide numerical evidence that in non-integrable models the dip of the entanglement revival decays faster than algebraically (likely exponentially).

\section{Free fermions}
\label{sec II}

In this section focus on the $XY$ chain that can be mapped to a free-fermion model. 
The $XY$ spin chain  with a transverse magnetic field has Hamiltonian 
\be
H=-\sum_{j=1}^L\left[ \frac{1+\gamma}{2}\hat S^x_j\hat S^x_{j+1}+ 
\frac{1-\gamma}{2}\hat S^y_j\hat S^y_{j+1}+ h \hat S^z_j\right],
\label{ti}
\ee
where $\hat S^{\alpha}_i$ are spin-$1/2$ operators acting at site $i$, $\gamma$ is 
the anisotropy, $h$ the transverse field, and we use periodic boundary condition.
The Hamiltonian ~\eqref{ti} can be diagonalized by a combination of Jordan-Wigner, Fourier transform,  and Bogoliubov 
transformations, leading to the free fermion model \cite{sach-book}
\begin{equation}
	H=\sum_{k} \epsilon_k \hat{c}_{k}^{\dag}\hat{c}_{k},\qquad\,\mathrm{with}\quad
\e_k^2=(h-\cos k)^2+\gamma^2\sin^2 k, 
\label{ti-1}
\end{equation}
and $\hat c_k$ are standard spinless fermionic ladder operators. 
The quasiparticle velocity is  $v(k)=d\e_{k}/dk$.
Note that the quasiparticles' velocities do not depend on the initial state because the 
system is non-interacting. However, the initial state determines 
the structure of the entangling quasiparticles produced after the quench, and hence the 
velocity at which entanglement spreads (see, for instance, Ref.~\onlinecite{btc-18} 
and~\onlinecite{najafi-2018}). 
Clearly, the velocities depend on the initial state in the 
presence of interactions~\cite{alba-2016,BEL-14}. 
 
Here we consider a quench of the magnetic field $h$ and of $\gamma$. 
Precisely, the system is initially prepared in the ground state of~\eqref{ti} with magnetic field $h_0$ and $\gamma _0$. 
Then, the parameters are suddenly changed as $h_0\to h$ and  $\gamma_0\to \gamma$.
The quench is parametrised in terms of the difference of the Bogoliubov angles of pre- and post-quench Hamiltonians, i.e. \cite{fagotti-2008,sps-04,calabrese-2005}
\begin{equation}
\label{eq:bol}
\cos \Delta_k=\frac{h h_0-\cos k (h+h_0)+\cos^2 k+\gamma\gamma_0\sin^2k}{\e\e_0},
\end{equation}
where $\e_0$, $\e$ stand for pre- and post-quench dispersion relations (see Eq.~\eqref{ti-1}).

The GGE built  with local conservation laws for the Hamiltonian ~\eqref{ti} is equivalent \cite{calabrese-2011,fe-13b} to the 
one built with mode occupation numbers $\hat{n}_k=\hat{c}_{k}^{\dag}\hat{c}_k$ and it is convenient to 
work with the latter. The GGE density matrix is then \cite{calabrese-2011,fe-13b} 
\begin{equation}
\rho_{\text{GGE}}=\frac{e^{-\sum_k \lambda_k \hat{n}_k}}{Z},
\end{equation}
where $Z=\Tr e^{-\sum_k \lambda_k \hat{n}_k}$ ensures the normalization $\Tr \rho _{\text{GGE}}=1$. 
The Lagrange multipliers $\lambda_k$ are fixed by imposing that the expectation 
value of $\hat{n}_k$ in the initial state coincides with its GGE prediction
\begin{eqnarray}
\langle \hat{n}_k\rangle _\text{GGE}=-\frac{\partial}{\partial \lambda_k}\sum _p \ln(1+e^{-\lambda_p}) =\frac{1}{1+e^{\lambda_k}}.
\end{eqnarray}
Now $\lambda_k$ is obtained requiring that $\langle \hat{n}_k\rangle _\text{GGE}=\langle \psi_0 | \hat{n}_k |\psi_0\rangle=n_k$. 
The thermodynamic entropy of the GGE is the thermodynamic entropy obtained from the occupation $n_k$, which reads 
\begin{eqnarray}
S_{\text{GGE}}=-\Tr [\rho_{\text{GGE}}\ln \rho_{\text{GGE}}] 
=\sum_k -n_k\ln n_k-(1-n_k) \ln (1-n_k) =\sum _k s(k),
\label{s(k)_tb}
\end{eqnarray}
where, $s(k)=-n_k\ln n_k-(1-n_k) \ln (1-n_k)$ is identified as the entropy contribution of the 
quasiparticle with momentum $k$.

In terms of~\eqref{eq:bol}, we have $n_k=\frac{1+\cos \Delta_k}{2}$ and so the quasiparticles entanglement content $s(k)$ reads 
\begin{equation}
	s(k)=-\frac{1+\cos \Delta_k}{2}\ln\Big(\frac{1+\cos \Delta_k}{2}\Big)-
	\frac{1-\cos \Delta_k}{2}\ln\Big(\frac{1-\cos \Delta_k}{2}\Big). 
\end{equation}
The quasiparticle prediction for the entanglement entropy for a quench in the XY chain is obtained by plugging this value of $s(k)$ and $v(k)=v_M\sin k $ in Eq.~\eqref{ee}. 
This time evolution (in the thermodynamic limit) has been also confirmed by ab initio approach both on the lattice \cite{fagotti-2008} 
and in the field theory limit \cite{clsv-19}.

\subsection{A special case: The N\'eel quench in the XX chain}
\label{sec IIa}

There is a special case of the quench above that we want to discuss separately because it will be important for the generalisation to 
interacting (both integrable and not) fermionic model. 
This is the time evolution starting from the Neel state (in the $z$ direction) and evolving with the isotropic XX Hamiltonian. 
This quench just amounts to take the limit $h_0\to-\infty$ and $\gamma\to 0$ in the above formulas, but we want to quickly discuss it here. 

For $\gamma=h=0$, the Jordan-Wigner transformation  maps the Hamiltonian to the tight-binding model 
\begin{eqnarray}
H=-\frac12 \sum_{i=1}^{L}\hat{c}_{i}^{\dag}\hat{c}_{i+1}+\text{h.c},
\label{tb}
\end{eqnarray}
where, $\hat{c}_i$ and $\hat{c}_{i}^{\dag}$ are local spinless fermionic annihilation and creation operators satisfying standard anti-commutation relations.   
The Hamiltonian~\eqref{tb} can be diagonalised in the momentum  basis and  can be written as
\begin{eqnarray}
H=\sum_{k} \epsilon_k \hat{c}_{k}^{\dag}\hat{c}_{k},
\label{tb1}
\end{eqnarray}
where, $\epsilon _k=-\cos k$ (the velocity is $v(k)=d\epsilon_k/dk=\sin k$ with maximum $v_M=1$ at $k_M=\pm \pi/2$). 
$\hat{c}_{k}^{\dag}$ and $\hat{c}_{k}$ are fermionic creation and annihilation operators in momentum space and they are closely  
related to the ones in Eq.~\eqref{ti-1} for the XY chain. 

The N\'eel state in fermionic basis is 
\begin{equation}
|\psi_0\rangle=\prod_{i=1}^{L/2}\hat{c}_{2i}^{\dag}|0\rangle.
\end{equation}
It is straightforward to check that for the N\'eel state in the thermodynamic limit $n_k=1/2$ for all $k$, which implies $s(k)=\ln 2$, 
as it can be also obtained from the limit of  Eq. \eqref{eq:bol}. 
This reflects the fact that all the nonzero overlaps between the N\'eel state and the eigenstates of the tight-binding chain are equal~\cite{mazza-2016}. 
The quasiparticle prediction for the entanglement entropy after the N\'eel quench is obtained by replacing $s(k)=\ln2$ and $v(k)=v_M\sin k $ in Eq.~\eqref{ee}. 

\begin{figure}[t]
\includegraphics[width=0.48\textwidth ] {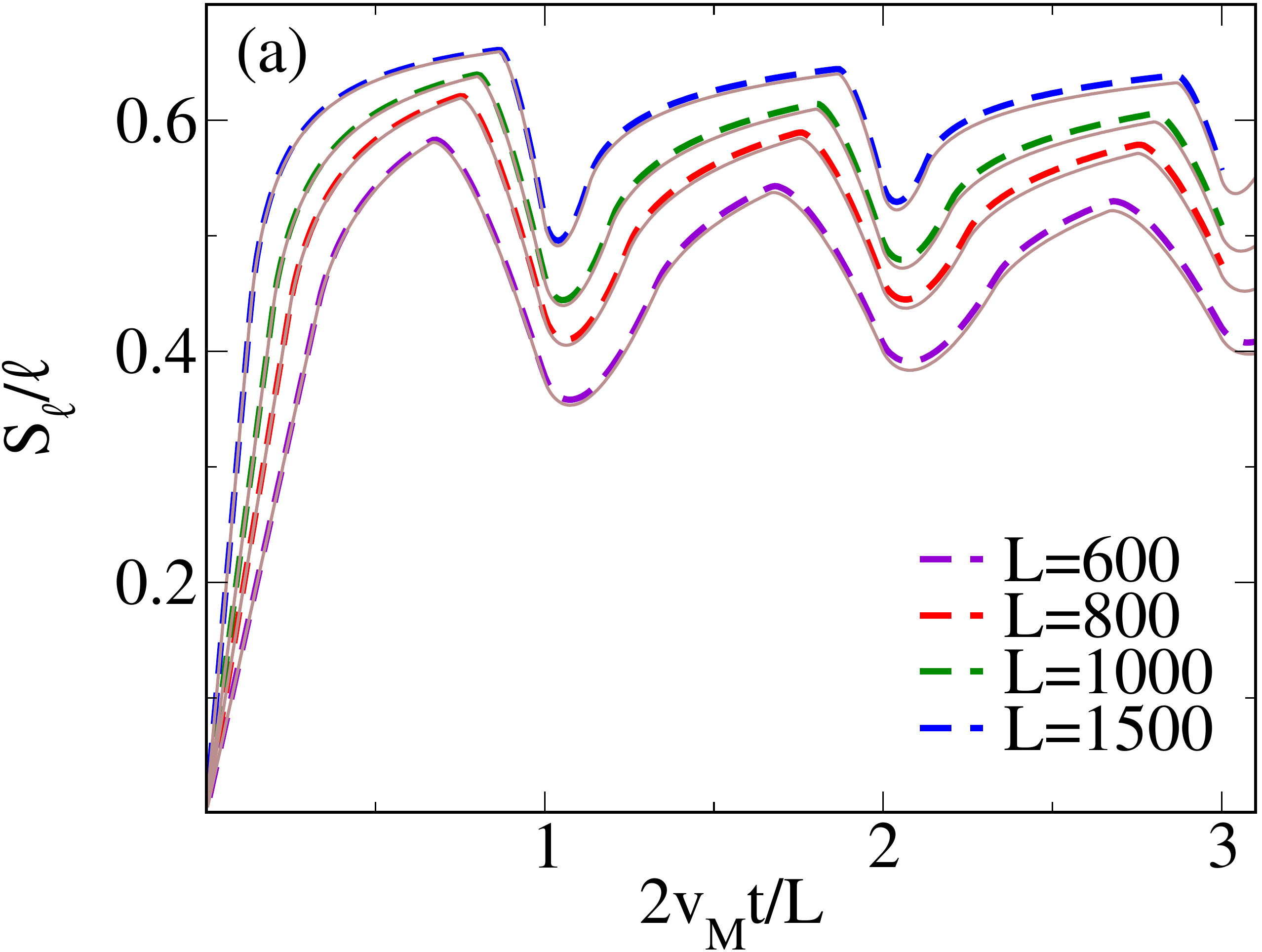}
\includegraphics[width=0.48\textwidth ] {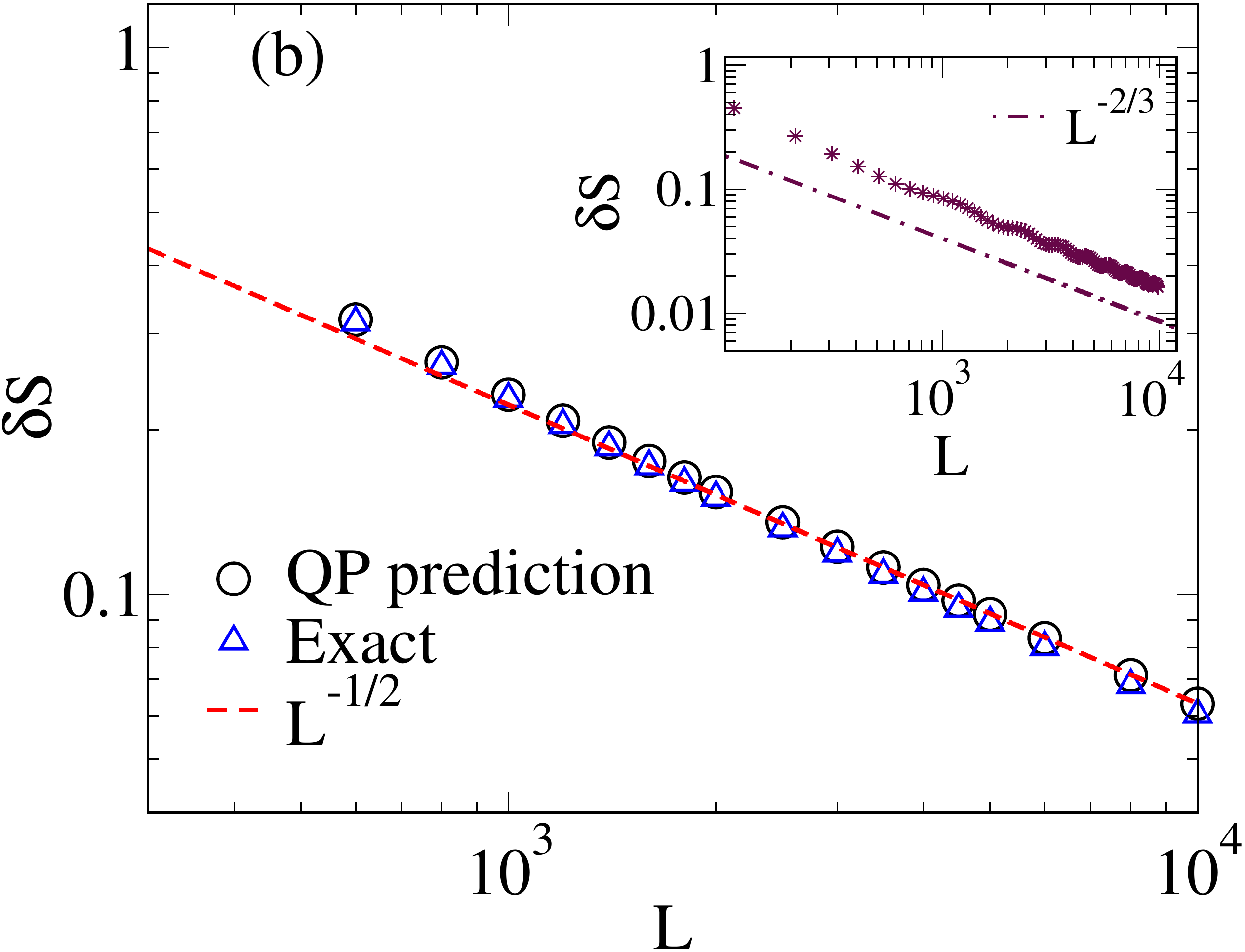}
\caption{(a) Entanglement revivals in the XX chain. Quench from the N\'eel state. $S_{\ell}/\ell$  plotted versus 
the rescaled time $2v_Mt/L$ for $L=600, 800, 1000, 1500$, and $\ell=200$. 
Solid  lines correspond to the quasiparticle prediction (see Eq.~\eqref{ee}), while dashed ones are exact numerical data. 
(b) Scaling of  the dip $\delta S$ of the first entanglement revival. 
In the main panel, we see that for $\ell=200$ as $L$ grows from $600$ to $10^4$ a scaling compatible with the quasiparticle prediction $L^{-1/2}$ is observed neatly.
However, for the larger values of $L$ a crossover towards a different scaling starts appearing. 
This crossover is highlighted in the inset, where we show that for $\ell=6$ the asymptotic scaling for large $L$ is $\sim L^{-2/3}$ (dot-dashed line).
}
\label{fig1}
\end{figure}

\subsection{Numerical study of the entanglement revivals}
\label{sec IIIa}

For free-fermion models the exact dynamics of the entanglement entropy is straightforwardly obtained from 
the time-dependent fermionic two-point correlations 
restricted to the subsystem $A$.
This technique is very standard and we do not review it here, the interested reader can consult the extensive literature on the subject \cite{peschel.2003,p-12,peschel.2009}. 

Let us first focus on the XX chain (the tight-binding model) after the quench from the N\'eel state. 
Our results are reported in Fig.~\ref{fig1}. 
Fig.~\ref{fig1} (a) shows the dynamics of the entanglement entropy $S_\ell$ for a subsystem 
of $\ell$ sites embedded in a chain of size $L$. We plot the entropy density $S_\ell/\ell$ 
for $\ell=200$ versus the rescaled time $2 v_Mt/L$, with $v_M$ the maximum velocity; the entanglement revival time is $t_R=L/(2v_M)$. The dashed lines are 
the exact numerical results for $L=600,800,1000,1500$. 
Several entanglement revivals are visible in the figure. 
The continuous lines  are the predictions using the quasiparticle picture  Eq.~\eqref{semi-cl}. 
They are in excellent agreement with the numerical results: in fact  $\ell=200$ is just a finite fraction of the total $L$ considered in the figure 
and we are safely working in the scaling limit $t,\ell,L\to\infty$ with fixed ratios.

Now, we  study how the height of the dip at the first entanglement revival time is damped as $L$ increases. 
To quantitatively address this behaviour, we  plot $\delta S$ versus $L$ in Fig.~\ref{fig1} (b). 
Here $\delta S$ is defined as in section~\ref{sec-I}, cf. Eq. \eqref{eq:delta-def}. 
We show $\delta S$ as a function of  $L$ for fixed  $\ell=200$.
It is evident that for the values of $L$ we considered (up to $10^4$) the quasiparticle picture works well,
since the numerical value for $\delta S$ and the quasiparticle one stay almost on top of each other. 
Still, it is evident that as $L$ increases, some deviations between the two appear and a decay faster than  $L^{-1/2}$ is starting. 
To highlight this crossover it is better to use smaller values of $\ell$ rather than increasing the one of $L$. 
Hence, in the inset of ~\ref{fig1} (b) we report again the scaling of $\delta S$ but for $\ell=6$ rather than $200$.
In it is clear that in this case the decay of $\delta S$ with $L$ is much faster and well described by the expected law $L^{-2/3}$. 
To summarise, Fig. \ref{fig1} confirms both the quasiparticle decays prediction~\eqref{ee_main} with $\delta S \sim L^{-1/2}$ at intermediate values of $L$
and the crossover to $L^{-2/3}$ for larger $L$. 
We recall that the origin of the exponent $2/3$ is related to the super-diffusion taking place on the light-cone of free fermion models related to Airy processes, 
see e.g. Refs. \cite{dsvc17,eisler-2013,clm-15,ms-16,vsdh-16,lms-18,stephan-2019} for similar results. 
For the quench from the N\'eel state in the $XX$ chain, the exponent $2/3$ can be derived analytically 
(see Appendix~\ref{sec:appendix}).

\begin{figure}[t]
\includegraphics[width=0.5\textwidth] {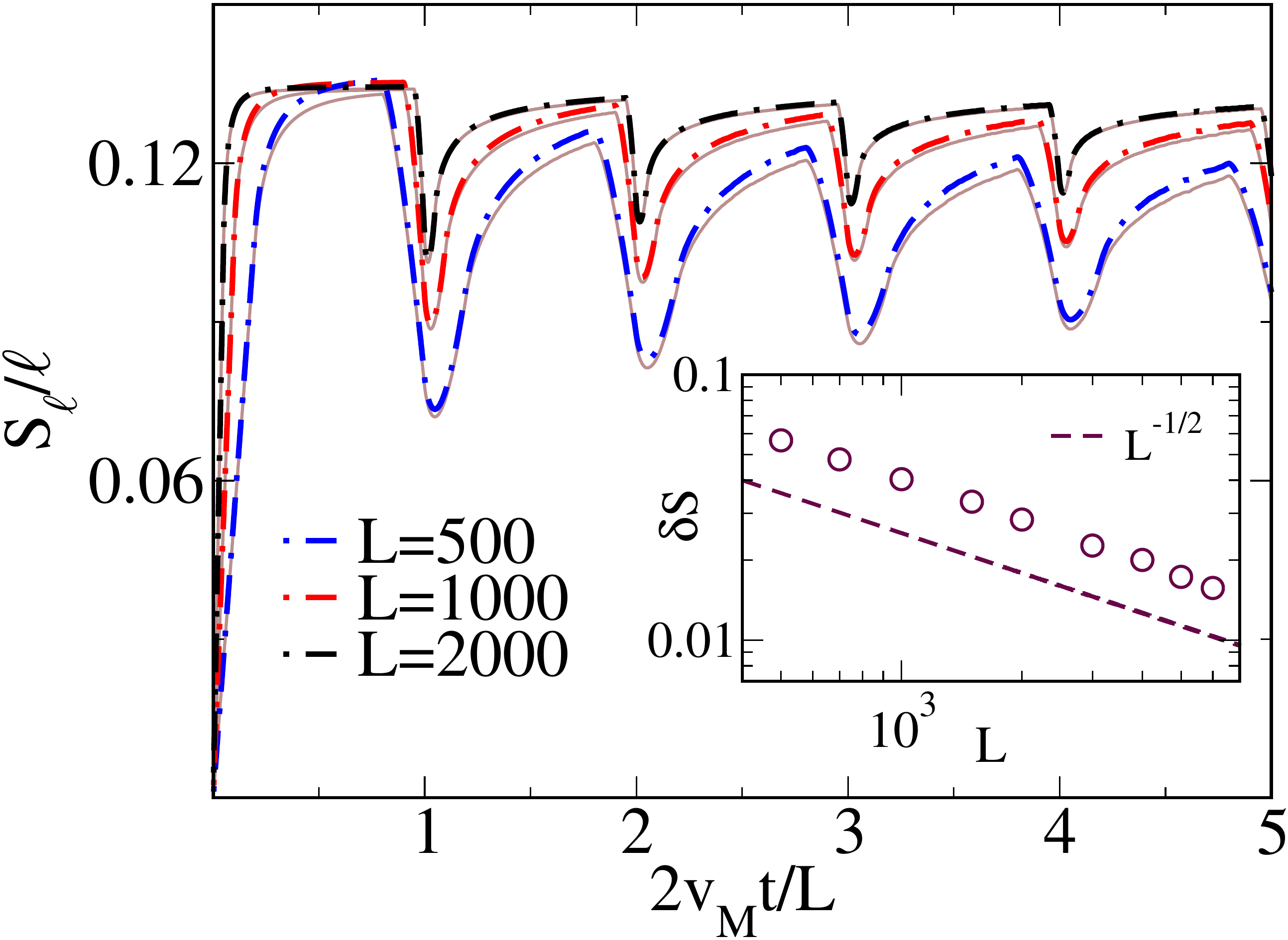}
\caption{Entanglement revivals in the $XY$ chain. $S_{\ell}/\ell$ versus rescaled 
time $2v_Mt/L$ for different values of $L$ after the quench from $(\gamma_0, h_0)$ to $(\gamma, h)$, with 
$\gamma=\gamma_0=1$ (Ising model), $h=2h_0=2.4$, and $\ell=100$.  
Solid lines correspond to  the quasiparticle prediction.
The inset shows the scaling of $\delta S$ with $L$ and the dashed line is a guide to the eye going as $L^{-1/2}$. }
\label{fig2}
\end{figure}

We end this section by briefly discussing numerical results for the quench in the $XY$ chain. 
We focus on the Ising case $\gamma=\gamma_0=1$.
The quench parameters for the transverse fields are fixed as $h=2, h_0=2.4$, but we checked that different values provide equivalent results. 
Figure~\ref{fig2} summarise our results for fixed $\ell=100$ and $L=500, 1000, 2000$. 
These values have been chosen to be sure to work within the scaling limit with large $\ell,L$ but with finite ratio. 
In the main panel we report $S_\ell/\ell$ as function of time: the agreement between the numerical data  and the quasiparticle prediction is always excellent. 
Curiously, the first entanglement revival $t_R=L/2v_M$ is quantitatively more pronounced than in the previously considered N\'eel quench in the XX chain. 
In the inset of Fig.~\ref{fig2} we fix again $\ell=100$ and study the scaling of the dip $\delta S$ as function of $L$ in the window $L\in[500,6000]$.
For these values of $L$, the numerical data are well captured by the quasiparticle  prediction  and scaling $\delta S \sim L^{-1/2}$. 
Also in this case, for the largest considered values of $L$ we observe a crossover towards a faster decay, that once again can be studied in more details. 
However, we do not perform this investigation here. 

\section{Free bosons} 
 \label{sec IIc}

In this section we  move our attention to  a bosonic system, namely the harmonic chain with Hamiltonian 
\begin{equation}
\label{hc-ham}
H=\frac{1}{2}\sum_{n=0}^{L-1} \left[ \pi_n^2+ m^2 \phi_n^2+ 
(\phi_{n+1}-\phi_n)^2\right], 
\end{equation}
with periodic boundary conditions. Eq.~\eqref{hc-ham} 
defines a chain of $L$ harmonic oscillators with frequency (mass) $m$ 
and with nearest-neighbor quadratic interactions. Here $\phi_n$ and 
$\pi_n$ are the position and the momentum operators of the $n$-th 
oscillator, with equal-time commutation relations $[\phi_m,\pi_n]=i\delta_{nm}$ and $[\phi_n,\phi_m]=[\pi_n,\pi_m]=0$.
Eq.~\eqref{hc-ham} is the lattice discretisation of a one-dimensional Klein-Gordon quantum field theory.

 The Hamiltonian ~\eqref{hc-ham} can be diagonalised in the momentum 
basis and it can be written as 
\begin{equation}
H=\sum_{k} \epsilon_k \hat{b}_{k}^{\dag}\hat{b}_{k}, \qquad
\label{ho1}
{\rm where }\quad
\epsilon _{k}^{2}=m^2+2\left(1-\cos k\right),
\end{equation}
and  $\hat{b}_{k}^{\dag}$ ($\hat{b}_{k}$) is the bosonic creation (annihilation) operator. 

Even for the harmonic chain, the GGE density matrix can be constructed from the mode occupation numbers $\hat{n}_k=\hat{b}_{k}^{\dag}\hat{b}_k$ \cite{cc-07,c-18}, 
and it is given by 
\begin{equation}
\rho_{\text{GGE}}=\frac{e^{-\sum_k \lambda_k \hat{n}_k}}{Z},
\end{equation}
where $Z=\Tr e^{-\sum_k \lambda_k \hat{n}_k}$ ensures the normalization $\Tr \rho _{\text{GGE}}=1$. 
The Lagrange multipliers $\lambda_k$ are fixed by imposing that the expectation value of $\hat{n}_k$ in the initial state coincides with its GGE prediction
\begin{eqnarray}
\langle \hat{n}_k\rangle _\text{GGE}=-\frac{\partial}{\partial \lambda_k}\sum _p \ln(1-e^{-\lambda_p}) =\frac{1}{e^{\lambda_k}-1}.
\end{eqnarray}
Now $\lambda_k$ can be obtained by imposing 
that the expectation value of $\hat{n}_k$ in the initial state $|\psi_0\rangle$ coincides with 
its GGE prediction, i.e., 
$\langle \hat{n}_k\rangle _\text{GGE}=\langle \psi_0 | \hat{n}_k |\psi_0\rangle=n_k$. 
The thermodynamic entropy is 
\begin{eqnarray}
S_{\text{GGE}}=-\Tr [\rho_{\text{GGE}}\ln \rho_{\text{GGE}}] 
=\sum_k  (1+n_k) \ln (1+n_k)-n_k\ln n_k =\sum _k s(k),
\label{s(k)}
\end{eqnarray}
where $s(k)=(1+n_k) \ln (1+n_k)-n_k\ln n_k$ is identified as the entropy contribution of the 
quasiparticle with momentum $k$. 


We consider the quantum quench in which the harmonic chain is 
initially prepared in the ground-state $|\psi_0\rangle$ of the Hamiltonian~\eqref{hc-ham} 
with $m=m_0$, and at time $t=0$ the mass is quenched to a different 
value $m\neq m_0$ ~\cite{cc-06,cc-07,sotiriadis-2014}. There have been extensive studies focusing 
on the critical ($m\to 0$)  and continuum limit ~\cite{nr-14, coser-2014, cotler-2016,sotiriadis-2016,bastianello-2016}. 
The validity of the quasiparticle picture for the entanglement entropy 
has been tested in all these situations \cite{alba.2018,nr-14, coser-2014, cotler-2016}. Note that for the quench to the massless case, 
a subleading logarithmic correction due to the presence of  zero mode appears \cite{cotler-2016}
and for this reason we will stay away from the critical case in the numerical analysis.

\begin{figure}[t]
\includegraphics[width=0.5\textwidth] {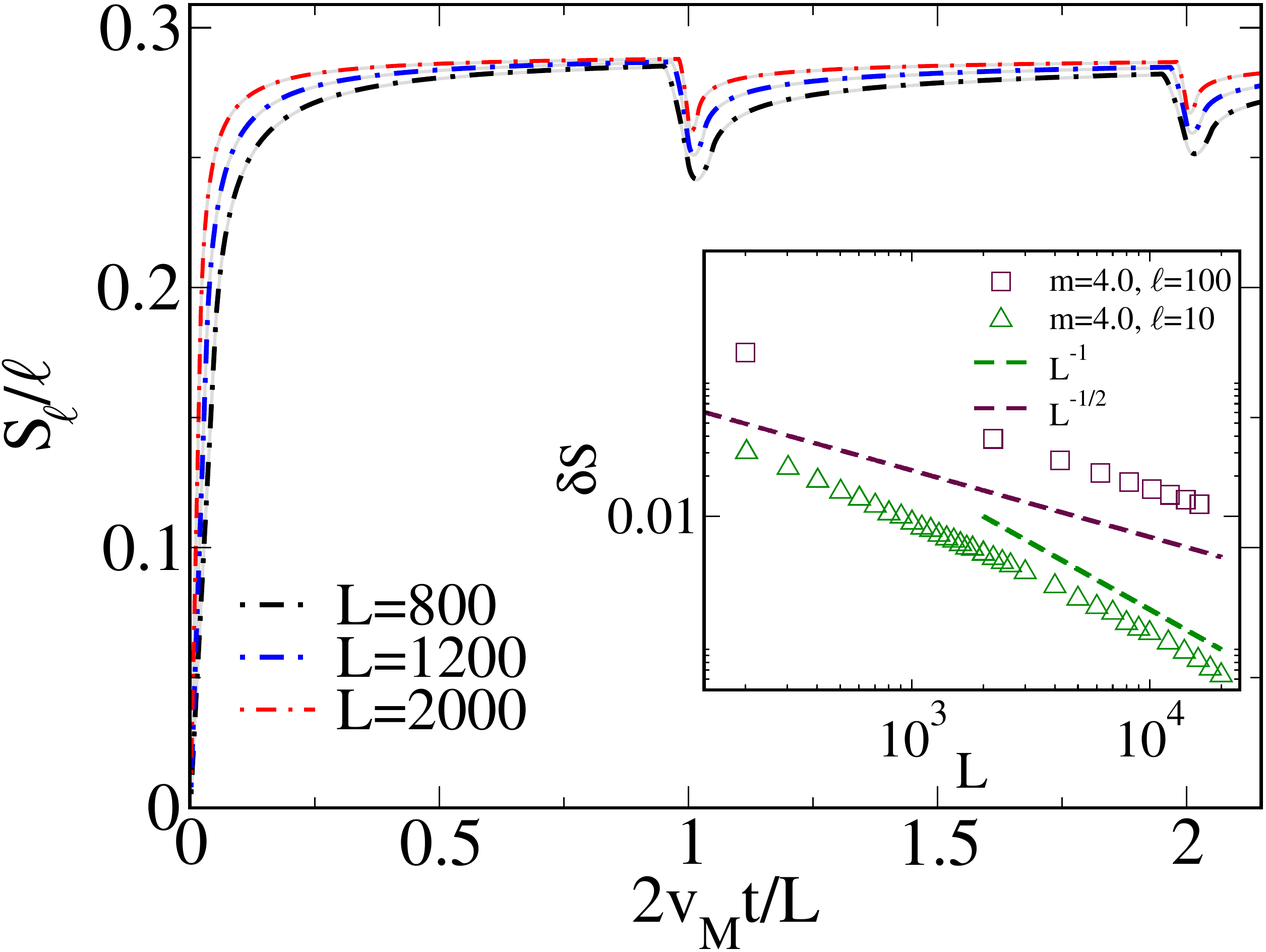}
\caption{ Entanglement revivals in the harmonic chain. $S_{\ell}/\ell$ versus rescaled time $2v_Mt/L$. 
 The data are for the  mass quench (from $m_0=2$  to  $m=4$), where $\ell=40$. Solid lines correspond to  the 
 quasiparticle prediction and dashed ones to the exact numerical data. 
 The inset shows the height of the entanglement revival $\delta S$  as a function of $L$ for two different 	quenches from $m_0=2$, both for $\ell=40$ and $\ell=10$. 
 While for $\ell=40$ the power-law  $\sim L^{-1/2}$ correctly captures our data, for $\ell=10$ there is a clear crossover to the truly asymptotic scaling $L^{-1}$. 
 }
\label{fig3}
\end{figure}

We use the notation $\epsilon^0_{k}$ for the 
dispersion relation in the initial state and $\epsilon_k$ for the one for $t>0$. 
The conserved occupation number $n_k$ for this quench is given by~\cite{alba.2018, cc-07,c-18}
\begin{eqnarray}
n_k&=&\langle \psi_0|\hat n_k|\psi_0\rangle=
\langle \psi_0 |a^\dag_k a_k|\psi_0\rangle=
\frac14\left(\frac{\epsilon_k}{\epsilon^0_{k}}+
\frac{\epsilon^0_{k}}{\epsilon_k}\right)-\frac12 \label{nkhc}\,.\label{vkhc}
\end{eqnarray}

We are now ready to  study the finite size behaviour of the entanglement entropy after a quench in the harmonic chain. 
The quasiparticle prediction for the entanglement entropy after this quench is given by Eq.~\eqref{ee}
in which we use $s(k)$ given by \eqref{s(k)} with \eqref{nkhc} and $v(k)= \frac{d\epsilon_k}{dk}$ with $\epsilon_k$ in \eqref{ho1}. 
Numerically, the entanglement dynamics is obtained (similarly to the case of free fermions) from the two-point correlation functions by use of 
standard techniques, as detailed e.g. in Refs.~\cite{peschel.1999,peschel.2009,coser-2014,p-12}. 
In Fig.~\ref{fig3} we report the numerical data for the time evolution of the entanglement entropy 
after the quench of the oscillators' mass from $m_0=2$ to $m=4$ (other values of the masses away from the critical point provide equivalent results),
for fixed $\ell=40$ and $L=800, 1200,2000$ chosen to be safely within the scaling limit of large $\ell,L$ with their ratio finite. 
The numerical data are compared  with the quasiparticle prediction~\eqref{ee} and the agreement is perfect. 

We now turn to the study of the dip at the first entanglement revival for the same quench parameters as above. 
The scaling of $\delta S$ as obtained from the numerical data is investigated in the inset of Fig.~\ref{fig3} as a function of $L$ (for fixed $\ell=40$).
As expected, since with these parameters we are working in the scaling limit,  we perfectly reproduce the scaling $\delta S\propto L^{-1/2}$.
However, also for bosons, for larger $L$ we expect a crossover to a different behaviour.
Employing the same logic as for fermions, rather than increasing the value of $L$, it is easier and more effective to reduce the one of $\ell$.
Hence, in the inset of  Fig.~\ref{fig3} we also report the data for $\ell=10$ that very clearly shows for large $L$ the crossover to the truly 
asymptotic behaviour $L^{-1}$. 
This scaling $L^{-1}$ is just a consequence of the fact that, away from the critical point, the non-equilibrium correlation functions of the  
harmonic chain are analytic and no singular scaling can take place.

\section{The Heisenberg $XXZ$ chain as a paradigm of integrable model}
\label{sec IId}

To investigate the effect of the interactions, here we consider the paradigm of integrable model, namely the 
spin-$1/2$ anisotropic Heisenberg chain (XXZ spin chain). The Hamiltonian is    
\begin{eqnarray}
 \hat H=\sum_{i=1}^{L-1} \left[ \frac{1}{2}\left( \hat S_{i}^{+} \hat S_{i+1}^{-} +\hat S_{i}^{-} \hat S_{i+1}^{+} \right) +  
 \Delta\Big(\hat S^{z}_{i}\hat S^{z}_{i+1}-\frac{1}{4}\Big)\right] .
 \label{def_ham1_tv}
\end{eqnarray}
Here $\hat S^{\alpha}_i$ are spin-$1/2$ operators acting at site $i$ of the chain and $\Delta$ is the anisotropy parameter. 
For $\Delta=0$, it reduces to the XX chain of section \ref{sec IIa} and it is mapped to free fermions. 
For $\Delta\neq 0$, the chain is genuinely interacting and  it can be solved by the Bethe ansatz \cite{takahashi,gaudin}.  

The non-equilibrium dynamics starting from many  initial states with low (mainly zero) entanglement has been considered in
several manuscripts in the literature \cite{fagotti-2013,p-13,fcec-14,ilievski-2015a,ilievski-2016b,brockmann.2014,wounters.2014,pozsgay.2014,ppv-17,mptw-15,ac-16qa,pvc-16,PVCR17,PPV-17}, 
employing different techniques  based on integrability.  
Here, for conciseness,  we  focus  on  the  non-equilibrium  dynamics of the entanglement entropy for a single quench, 
although we expect our results to be more general.  
Namely, we only consider the dynamics starting  from (one of the two degenerate) N\'eel state 
\begin{equation}
	|\psi_0\rangle=
	\left|\uparrow\downarrow\uparrow\dots\right\rangle.
\end{equation}
The GGE describing the steady state after the quench from the N\'eel state can be constructed analytically using the Quench-Action 
method~\cite{brockmann.2014,wounters.2014}. 
Based on this solution for the Bethe ansatz root density, and exploiting Eq. \eqref{semi-cl}, the quasiparticle prediction for the entanglement dynamics after the quench from  
the N\'eel state in the thermodynamic limit has been explicitly written down in Refs.~\cite{alba-2016,alba.2018} for $\Delta\ge 1$. 
In Eq. \eqref{semi-cl} one just needs to interpret $s_n(\lambda)$ as the Yang-Yang entropy densities of the GGE and $v_n(\lambda)$ as the 
velocities of the excitations built on top of the excited state itself. 

\begin{figure}
\includegraphics[width=0.49\textwidth] {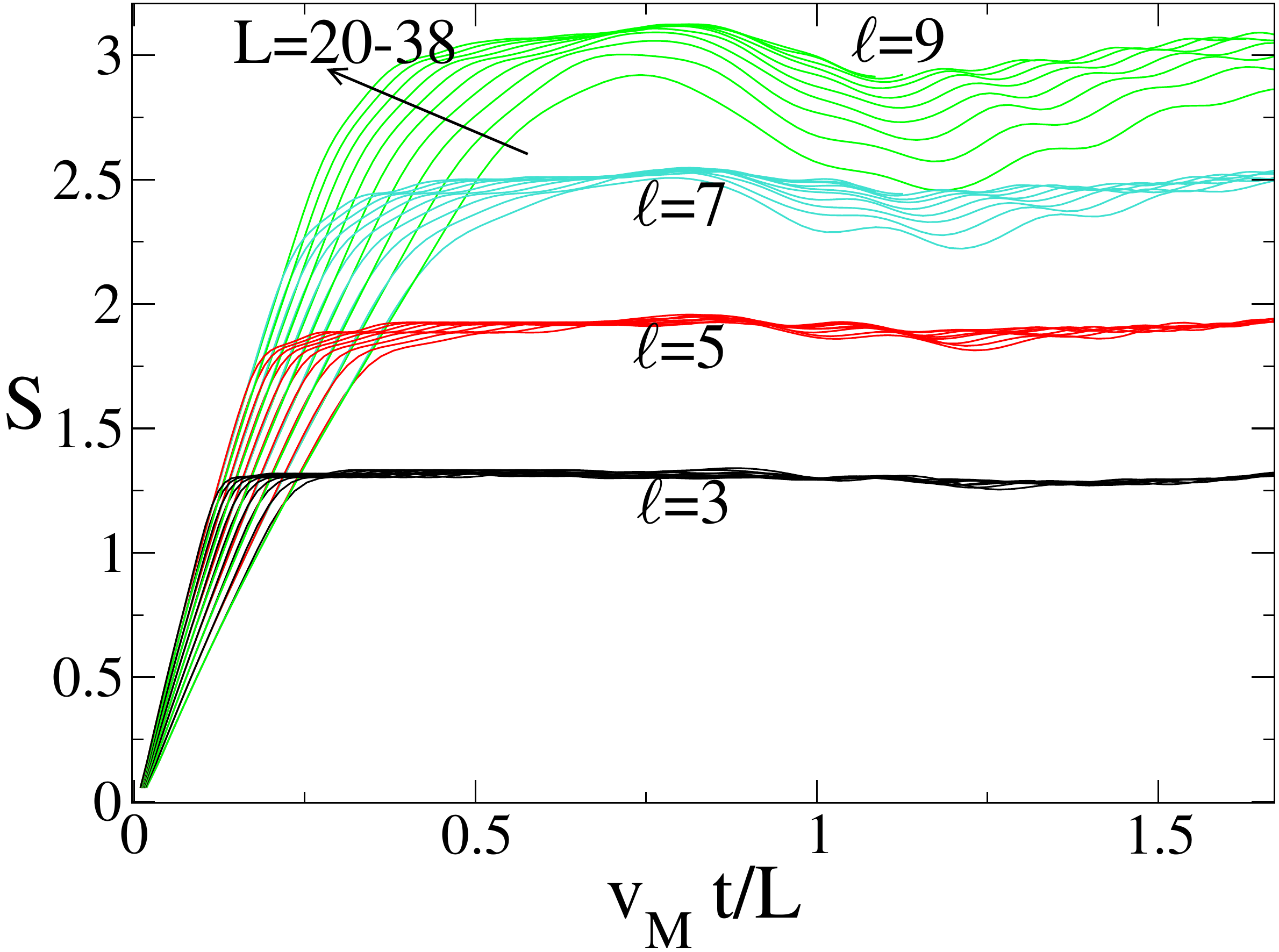}
\caption{Time evolution of the entanglement entropy after a quench from the N\'eel state to the XXZ Hamiltonian with $\Delta=2$ and with open boundary conditions.
Four values of $\ell=3,5,7,9$ are displayed and $L$ ranges in the interval $L\in[20,38]$. 
Revivals are evident for the larger considered $\ell$, but their quantitative analysis is made difficult by the presence of spurious 
finite-size effects. Although we do not report the analysis  here, the dip $\delta S$ at $\ell=9$ is algebraic with exponent between $1$ and $2$. 
}
\label{fig_D2}
\end{figure}

However, despite the integrability of the model, calculating the entanglement entropy, even at equilibrium \cite{atc-09}, is a daunting task.
Furthermore, the study of the time evolution exploiting integrability is extremely difficult  even for simpler observables \cite{NaPC15,dc-14}. 
Hence, in order to access the non-equilibrium dynamics of the entanglement entropy we have to resort to numerical simulations.
Here we use tDMRG techniques~\cite{white-2004, daley-2004, uli-2011}. 
Consequently, all the data for entanglement dynamics reported in this and in the following Section are obtained using the tDMRG algorithm, 
as  implemented  in  the  iTensor  library ~\cite{itensor}. 
We  work with  the  maximum  bond  dimension  $\chi=2000$ and with a time step  $\Delta t=0.02$ for all simulations. 
We have checked the robustness of our results by performing simulations for different values of $\delta t$ and $\chi$. 
Furthermore, all of our results have been verified with exact diagonalisation for small system sizes.

The only minor drawback of tDMRG is that it works more effectively with open boundary conditions, rather than periodic ones. 
This is not a main problem for the entanglement revivals dynamics:  
they are always present in the entanglement dynamics but now they are due to quasiparticles bouncing back at 
the chain boundaries, rather than going around the full chain. 
This effect amounts to a minor modification in the semiclassical quasiparticles' trajectories so that 
the entanglement revival occurs at $t_R=L/v_M$ and not at $t=L/(2v_M)$ as for a periodic chain.

At this point it is rather natural to start our analysis from the case $\Delta\geq1$ for which an exact solution for the stationary state exists.
This implies that we also know exactly the value of the maximal velocity and the stationary entropy, both important for the study of the dip 
of the entanglement revivals. 
The numerical time evolution of the entanglement entropy after a quench from the N\'eel state to the XXZ Hamiltonian is reported in Fig. \ref{fig_D2}
for $\Delta=2$ (different values for all $\Delta \geq 1$ produces qualitatively equivalent results).
With tDMRG we can access relatively small values of $\ell$ and $L$. 
In the figure we report $\ell=3,5,7,9$ (the data for even $\ell$ would be too close and make the figure difficult to read, but they are very similar) 
and we focus on $L$ in the interval $L\in[20,38]$. 
With a large numerical effort it is possible to increase slightly these values, but for a more quantitative analysis, one needs to at least double both $\ell$ and $L$.
In the figure, entanglement revivals are evident for $\ell=7$ and $\ell=9$, although they take place slightly after $L/v_M$.
This is clearly due to the presence of many species of particles (called strings) which are not yet well separated for $\ell=9$.
The resulting finite size effects are huge and a quantitative analysis of the dip of the entanglement revival is unstable, although 
we know a priori the values of $t_R$ and $S_\ell(\infty)$ that enter in the definition \eqref{eq:delta-def} of $\delta S$.
Although we do not present any analysis of the dip of the first entanglement revival, the data at both $\ell=9$ and $\ell=7$ are well compatible with 
an algebraic decay of $\delta S$ with an exponent which is between $1$ and $2$. 
This is different from the quasiparticle prediction $1/2$, but we cannot conclude whether this is due to the oscillations of the data or 
to the fact that we are not yet in the scaling regime, although we tend to believe more to the second explanation.

\begin{figure}
\includegraphics[width=0.49\textwidth] {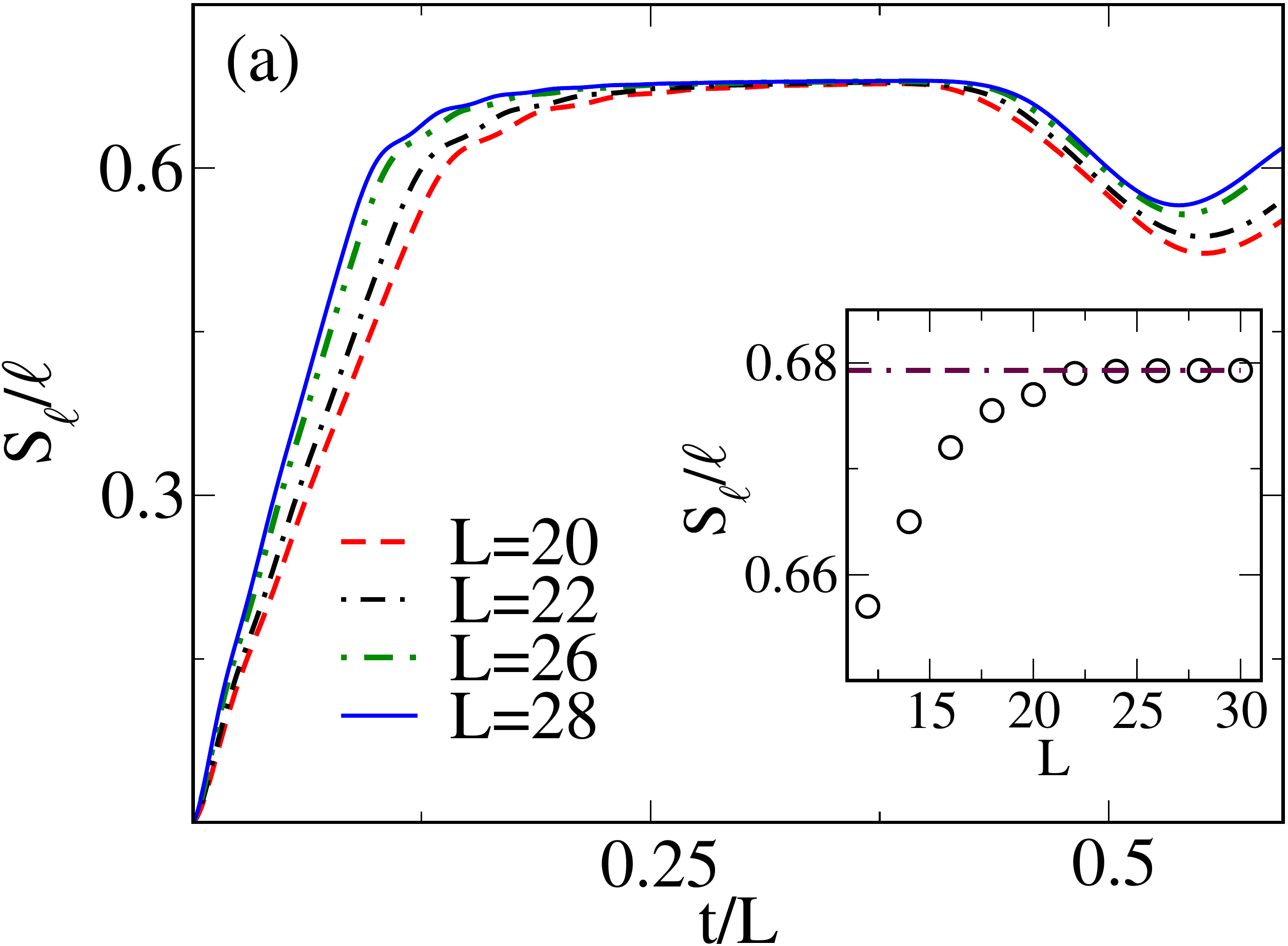}
\includegraphics[width=0.49\textwidth] {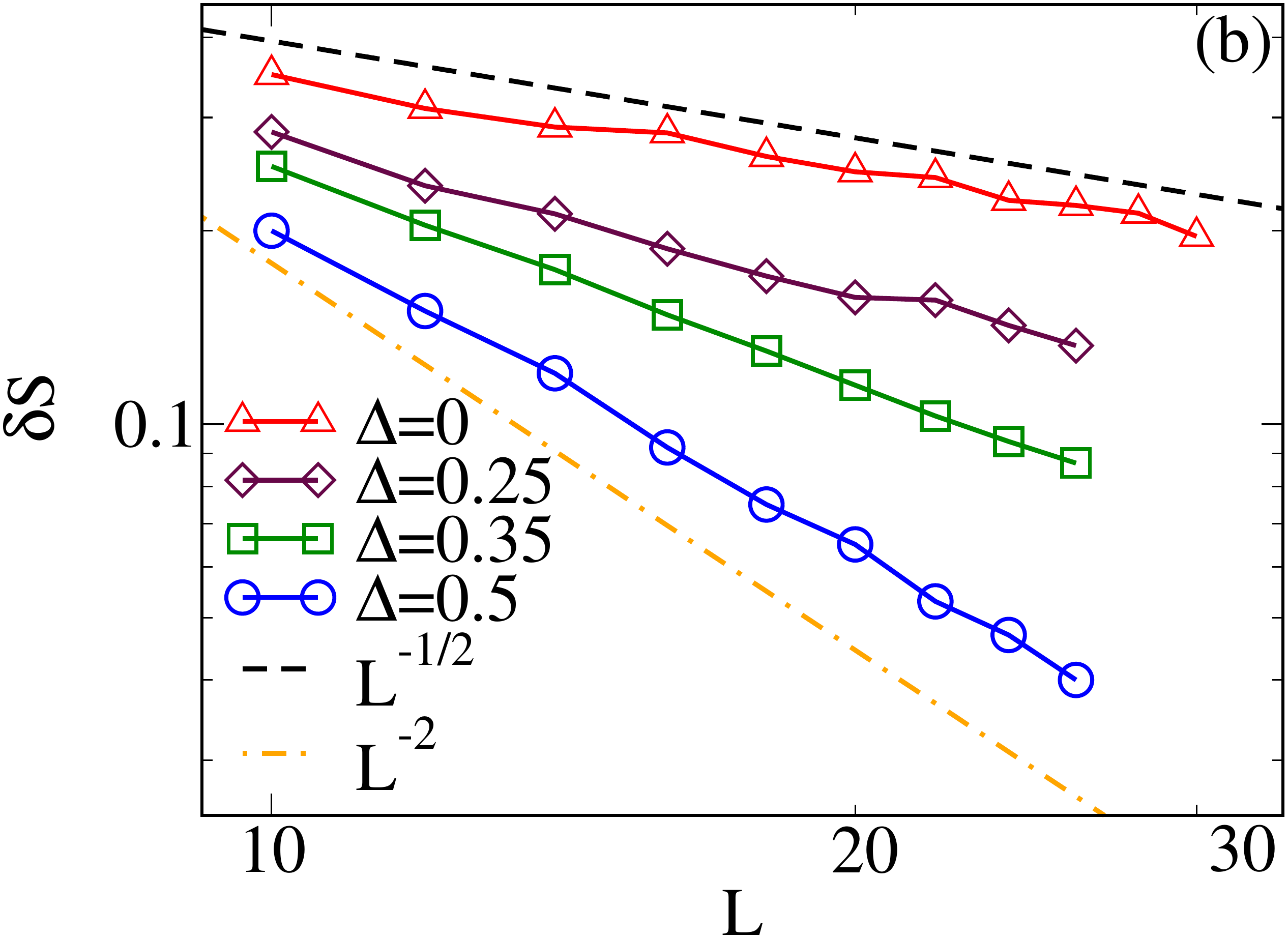}
\caption{Entanglement revivals in the Heisenberg chain for $\Delta<1$ with tDMRG simulations for open boundary conditions.  
(a) $S_{\ell}/\ell$  versus  rescaled time $t/L$ for $\Delta=0.25$.  
Several values of $L$ are shown. The value of $\ell$ is fixed to $\ell=4$. 
The inset shows the density of entanglement entropy  $S_\ell/\ell$ in the plateau as a function of $L$. 
The dashed line is the extracted value for the steady state entropy. 
(b) The scaling of the dip $\delta S$ as a function of $L$ for $\Delta=0,0.25,0.35,0.5$. 
The dashed (dot-dashed) line is a guide to the eye going like $\sim L^{-1/2}$ ($L^{-2}$). 
}
\label{fig4}
\end{figure}

Since the data for $\Delta>1$ are quantitatively not  convincing enough, it is rather natural to move our attention to the window $\Delta\in[0,1)$. 
The main advantage is that we expect smaller finite size effects because (at rational values of $\Delta$) the number of species of quasiparticles is  
finite (often small) and so they should produce less interference effects and clearer entanglement revivals. 
The steady-state after the quench from the N\'eel state can be in principle obtained by using the approach 
of Ref.~\onlinecite{ilievski-2016b} (see also~\onlinecite{fagotti-2013,deluca-2017}). 
Hence, we do not know exactly either the value of $S_\ell(\infty)$ or that of $t_R$ (because we do not know $v_M$).
The time evolution of the entanglement entropy for $\Delta=0.25$ is reported in Figure~\ref{fig4}(a).  
We focus on $\ell=4$ and $L$ in the range $L\in[20,28]$.
The reason of this small value of $\ell$ (compared to the study at $\Delta>1$ in Fig. \ref{fig_D2}) is that the entanglement entropy grows 
much faster (its asymptotic density is more than the double than at $\Delta=2$), limiting the performance of the tDMRG algorithm. 
However, even at this small value of $\ell$ we see a neat first entanglement revival with very small finite size effects, confirming our expectations. 
In this case, we do not know the asymptotic value of the entropy, but we can extract it in a very robust manner from the 
numerics since the plateau is very stable (as a difference with the case $\Delta>1$). 
The determination $S_4(\infty)$ is reported for $\Delta=0.25$ in the inset of Fig.  \ref{fig4}(a).
Curiously enough, the asymptotic value $S_\ell(\infty)$ depends very little on $\Delta$ in all the window $\Delta\in[0,1)$ and it is very close to 
the value at $\Delta=0$, i.e. $\ell \ln 2$ (but it is definitively different). 
In order to study the behaviour of the dip of $\delta S$ as function of $L$ (see again Fig.~\ref{fig0} (a)) for fixed $\ell$, 
we would need to know $v_M$ to extract $t_R$. To circumvent this problem, we instead use in Eq. \eqref{eq:delta-def} the very neat minimum that 
the entanglement entropy displays. The difference between $S_\ell(t_R)$ and the minimum is anyhow expected to shrink and go to zero for large $L$;
so, in the worst case scenario, this replacement introduces just a further finite size effect. 
Calculating $\delta S$ in this way, we report our data for several $\Delta\in[0,1]$ in Fig.~\ref{fig4} (b), all for $\ell=4$. 
Since the plot is in log-log scale, the data are all fully compatible with a power-law behaviour with an exponent between  $1/2$ and $2$
(these two extremes are shown as guides to the eyes in the plots).   
We could also extract effective exponents for the decay, but we do not find their values significative for these small value of $\ell$ and $L$.  

We stress that upon increasing the values of $\ell$ and $L$, we expect the data to crossover to the quasiparticle picture results
with a behaviour of the dip going like $L^{-1/2}$, also in the presence of interactions for any $\Delta$. 
However, unlike for free models, here we are restricted to the system size $L\simeq 30-40$ due to the limitation of available  numerical techniques. 
Hence, the quasiparticle regime cannot be accessed. 
In fact,  the prediction for the time evolution from the quasiparticle picture in Fig. \ref{fig_D2} together with the numerical data for $\Delta>1$, 
the agreement would be satisfactory only at short time, as already shown in Refs. \cite{alba-2016,alba.2018}.
Anyhow, in spite of the many drawbacks of this quantitative analysis, we can conclude, without doubts, that the decay of the dip of the entanglement revival 
is algebraic with an exponent between $1/2$ and   $2$ for all values of $\Delta$. 
Such a conclusion also reenforces the same result found for $\Delta>1$ but with less stable data.

\section{Interactions that break integrability}
\label{sec IIIc}

We have verified in the previous sections the validity of the quasiparticle picture 
Eq.~\eqref{ee} to describe the entanglement revivals in free fermionic and 
free bosonic models. Moreover, we showed that the dip in the 
entanglement entropy at the entanglement revival time is damped as $L^{-1/2}$ in the scaling regime (cf.~\eqref{ee_main}) for 
free-fermion and free-boson models. 
For interacting integrable models it is difficult to access the scaling regime, but we can anyhow show that the decays of the dip
is algebraic with $L$.
We now discuss the effects of integrability  breaking. 

As an example of a chaotic model, we consider a non-integrable perturbation added to the XXZ Hamiltonian \eqref{def_ham1_tv}, in such a way to have 
an integrable limit. 
Among the various possibilities to break integrability, we opt for the addition of a longitudinal magnetic field $h_x$ modifying the Hamiltonian as
\begin{eqnarray}
 \hat H=\sum_{i=1}^{L-1} \frac{1}{2}\left( \hat S_{i}^{+} \hat S_{i+1}^{-} +\hat S_{i}^{-} \hat S_{i+1}^{+} \right) +  
 \Delta\left(\hat S^{z}_{i}\hat S^{z}_{i+1}-\frac{1}{4}\right) +h_x\sum_{i}\hat{S}^{x}_{i}.
 \label{def_ham1_tv2}
\end{eqnarray}
The Hamiltonian ~\eqref{def_ham1_tv} is non-integrable except at the isotropic point $\Delta=1$. 

\begin{figure}[t]
\includegraphics[width=0.473\textwidth] {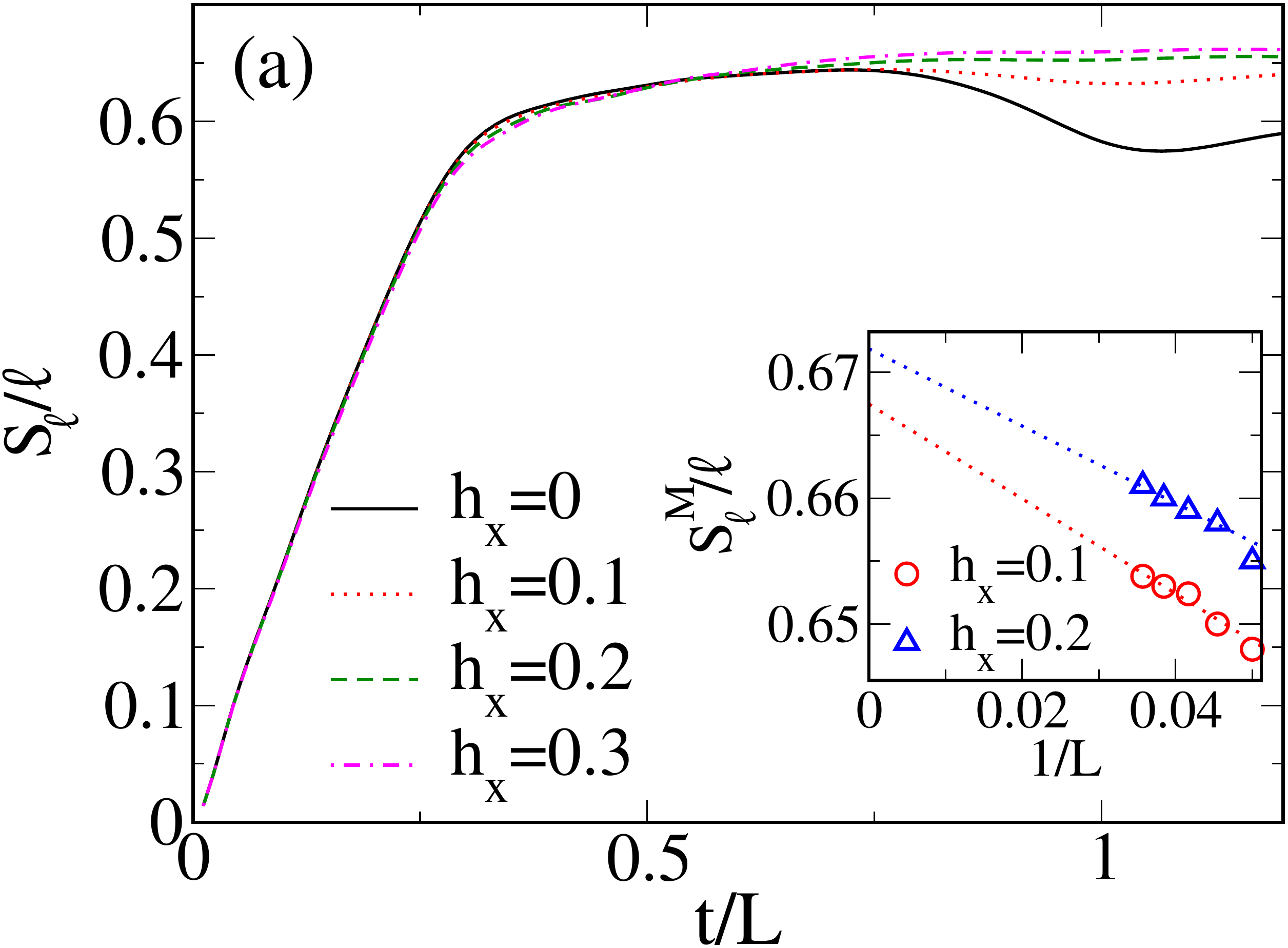}
\includegraphics[width=0.51\textwidth] {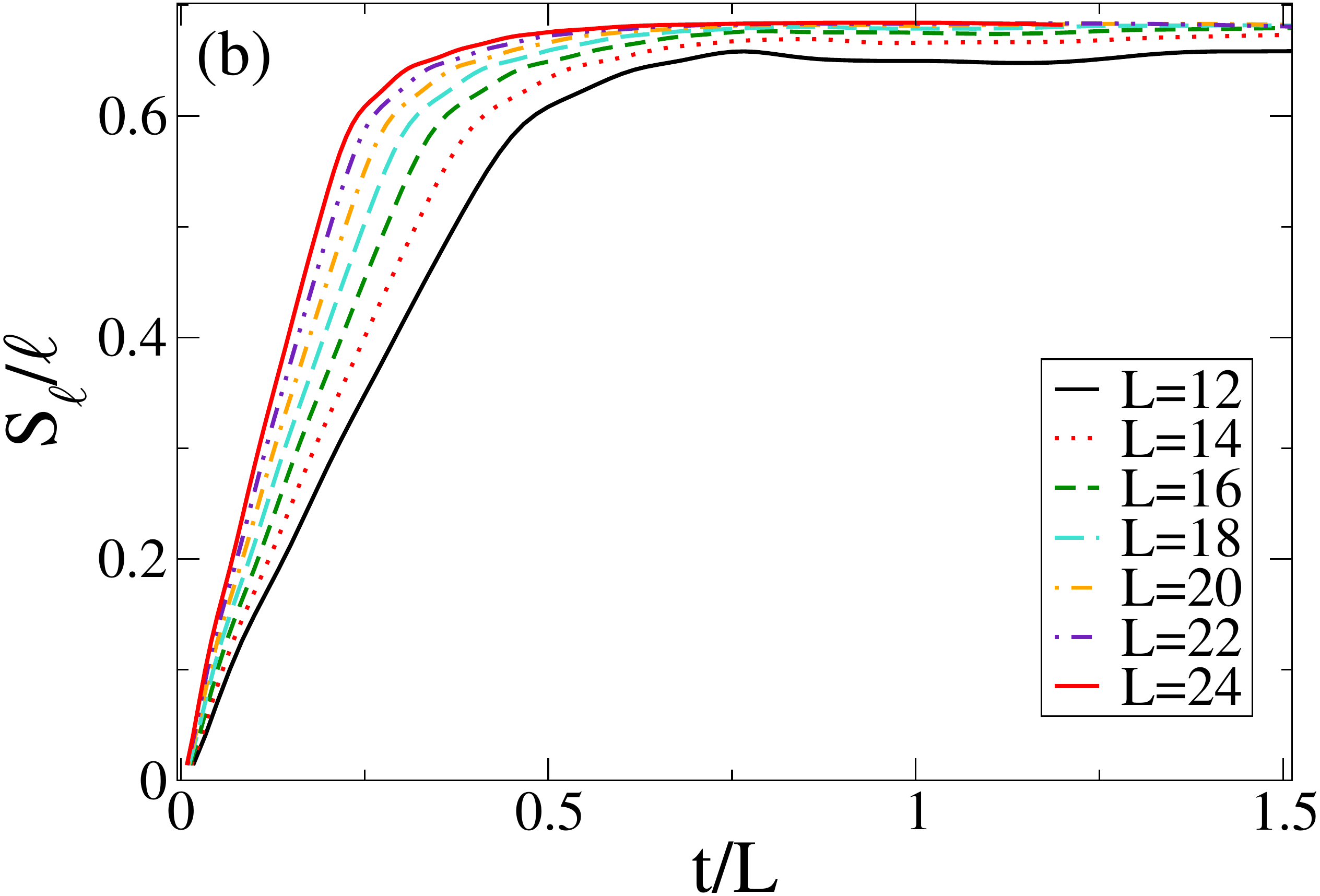}
\caption{Entanglement revivals in the non-integrable Heisenberg chain \eqref{def_ham1_tv2} with open boundary conditions after a quench from the N\'eel state. 
(a) Entropy density $S_{\ell}/\ell$ for $L=18$ and $\ell=4$ versus  rescaled  time $t/L$. 
Here $\Delta=0.5$ and  $h_x=0, 0.1, 0.2, 0.3$. 
The inset shows the maximum of $S_{\ell}/\ell$ versus  $L$ for $h_x=0.1,0.2$. 
Dashed lines corresponds to the best fit to $S^M_\ell/\ell=S_\ell(\infty)/\ell+a/L$ with $S_\ell(\infty)$ and $a$ fitting parameters. 
(b) The same as in (a) but at fixed $h_x=0.2$ with varying $L$.
}
\label{fig5}
\end{figure}

As for the integrable XXZ chain, we use tDMRG simulation to access the entanglement dynamics following a quench governed by the Hamiltonian  \eqref{def_ham1_tv2}.
Again we focus on the dynamics starting from the N\'eel state and we only consider $\Delta=0.5$ which is far enough from the integrable point at $\Delta=1$. 
Figure~\ref{fig5} shows the entanglement evolution  for different values of $h_x$ for  fixed $\ell=4$ and various $L$ (only (b) panel). 
It is clear from Fig.~\ref{fig5}(a) that, at fixed $L$, the entanglement revival fades away as the integrability breaking parameter $h_x$ is increased. 
However, to perform a quantitative analysis of the dip $\delta S$, we  need first to numerically extract $S_\ell(\infty)$.  
Unlike the integrable case (see Fig.~\ref{fig4}), here we observe clear finite-size effects. 
In the inset of Fig.~\ref{fig5}(a), we  show the maximum of $S_\ell/\ell$  as a function of  $1/L$. 
The data suggest to fit $S_\ell^M/\ell$ with a linear form $S_\ell^M/\ell=S_\ell(\infty)/\ell+a/L$  (where $S_\ell(\infty)$ and 
$a$ are fitting parameters). Interestingly, we observe that $S_\ell(\infty)/\ell\approx 0.67$, 
which is close to the maximum entropy density $\ln2$, which is also the value at $\Delta=h_x=0$. 
Since the model is not integrable, $t_R$ is not well defined, so in the definition of $\delta S$ we can use the minimum of $S_\ell$ close to the entanglement revival, 
as we did for the integrable XXZ spin-chain. 
Anyhow it is very difficult to evaluate this minimum for $h_x\neq 0$ as soon as $L$ is moderately large. 
For example in Fig.~\ref{fig5} (b) we report the time evolution of the entanglement entropy at fixed $h_x=0.2$, but for different values of $L$. 
We can observe a small entanglement revival only for $L=12$, but for larger $L$ the decay is too fast to identify a clear minimum in the entropy. 
If one attempts an analysis at this not well-identified dip, the decay is much faster that any power-law, likely exponential, but could be even faster. 
Our findings  match well the results found in Ref.~\cite{alba.2019} from the scaling of the peak of the mutual information. This is not surprising because 
the physical origin of the mutual information scrambling and of the decay 
of the entanglement revivals is the same, i.e. that in 
non-integrable systems there are no well defined quasiparticles which can preserve 
quantum correlations over large distance and long times. 
We should mention that the random unitary~\cite{nahum-17} framework predicts the 
absence of revival, as it is observed in Ref.~\onlinecite{bkp-20}.

\section{Conclusions}
\label{sec IV}

In this work we investigated the revivals of the entanglement entropy of an interval of fixed length after a quantum quench in finite-size systems. 
Our main result is that both in integrable and in non-integrable systems the strength of the entanglement revival is damped as a function of the system size. 
However, while the damping is algebraic for 
integrable systems, it is much faster in chaotic ones (strong scrambling). 
Within the quasiparticle picture for the entanglement spreading of integrable models, the exponent of the power-law decay is $1/2$. 
However, we provide compelling evidence that this exponent describes only an intermediate regime for $\ell=a L$ with $a\ll 1$.
For very small $\ell$, there is a crossover toward a truly asymptotic regime with a model dependent exponent which is larger than $1/2$ 
(e.g., it is equal to $2/3$ for the XX chain and to $1$ for gapped free bosons).  
Our results suggest that the entanglement revivals provide a useful tool to diagnose scrambling in quantum many-body systems.

There are several interesting directions for future work. 
An simple generalisation of our work would be to understand the scaling of entanglement revivals in higher-dimensional free models, for 
which the same techniques used here trivially apply. 
A more difficult generalisation would be to understand what happens for R\'enyi entropies.
Indeed, while for integrable models the quasiparticle picture is expected to work (with some troubles though, see Refs.
\cite{alba_renyi_qa.2017,alba_renyi.2017,alba_renyi.2019,mestyan.2018}), for chaotic systems with a conserved charge 
there are effects of diffusion also at intermediate times \cite{zl-20,z-20,rpk-18,kvh-18} and it is unclear how they could affect the entnaglement revivals. 
Another natural question concerns whether entanglement revivals can survive to the addition of some quenched disorder and, if yes, how 
they disappear as the system size grows. 
Finally, in this work we focused on entanglement revivals after a global quantum quench. However, 
entanglement revivals are expected also in different non-equilibrium protocols such as, just to quote two examples, the domain-wall quench~\cite{collura.2018,misguich.2017,cdcd-20}
and geometric quenches~\cite{alba-2014,ge-19}. It would be interesting to understand the damping of entanglement revivals also in these situations. 

Finally, it has been suggested recently that the eigenstate-average of the 
entanglement entropy can potentially distinguish between integrable and chaotic systems~\cite{vidmar-2019}. Specifically, the prefactor of the volume-law entropy seems to exhibit a marked 
difference between integrable and chaotic dynamics, and it depends in  a nontrivial 
fashion on the aspect ratio $\ell/L$. It would be interesting to device an out-of-equilibrium 
protocol that allows to reveal this behavior, for instance considering Floquet 
systems. This could provide a further out-of-equilibrium diagnostic tool to distinguish 
integrable versus chaotic models. 

\section{Acknowledgments}
VA~acknowledges support from the European Research Council under ERC Advanced grant 743032 DYNAMINT.
PC acknowledges support from ERC under Consolidator grant number 771536 (NEMO).

\appendix

\section{Decay of entanglement revival in the $XX$ chain after the N\'eel quench}
\label{sec:appendix}

Here we derive the exponent $2/3$ of the decay of the entanglement 
revival after the quench from the N\'eel state in the $XX$ chain 
(see section~\ref{sec IIIa}). 

A straightforward application of Wick's theorem gives the 
correlation matrix $C_{nm}\equiv\langle\mathrm{Neel}|c^\dagger(t)c(t)
|\mathrm{Neel}\rangle$ after the N\'eel quench as 
\begin{equation}
	\label{eq:app-1}
	C_{nm}=\frac{1}{2}\delta_{nm}+\frac{(-1)^m}{2}\frac{1}{L}
	\sum_{k=-(L-1)/2}^{(L-1)/2}e^{-i 2\pi k/L(n-m)+2it\cos(2\pi k/L)}
\end{equation}
One has to derive the asymptotic behavior of $C_{nm}$ in 
the limit $L\to\infty$. This can be done in  a standard way 
by using the Poisson summation formula. For a generic periodic 
function $G(x)=G(x+2\pi)$ this states that 
\begin{equation}
	\frac{1}{L}\sum_{k=-(L-1)/2}^{(L-1)/2}G(2\pi k/L)=\frac{1}{2\pi}\sum_{l=-\infty}^\infty
	\int_{-\pi}^\pi dq G(q)e^{i q l L}. 
\end{equation}
Now, Eq.~\eqref{eq:app-1} gives 
\begin{equation}
	\label{eq:int}
	C_{nm}=\frac{1}{2}\delta_{nm} +\frac{(-1)^m}{2}\frac{1}{2\pi}\sum_{l=-\infty}^\infty
	\int_{-\pi}^\pi dq e^{-i q(l L -n+m)+2it\cos q}
\end{equation}
The integral above can be performed analytically to obtain 
\begin{equation}
	C_{nm}=\frac{1}{2}\delta_{nm} +\frac{(-1)^m}{2}\sum_{l=-\infty}^\infty 
	i^{l L-n+m}J_{l L-n+m}(2t), 
\end{equation}
where $J_p(x)$ is the Bessel function of the first kind. 

To proceed we consider the correlation function at the 
revival time $t=L/2$. 
Thus, a straightforward analysis of the asymptotic behavior of the 
Bessel functions in the limit $L\to\infty$, or, equivalently 
a saddle point treatment of the integral in~\eqref{eq:int} shows 
that the terms with $l\ne-1,0,1$ give exponentially suppressed 
contributions in the limit $L\to\infty$. The remaining terms give 
\begin{equation}
	\label{eq:c-fin}
	C_{nm}=\frac{1}{2}\delta_{nm} +
	\frac{6^\frac{1}{3} (-1)^{m+1} \cos \left(\frac{1}{2} \pi  (n-m)\right)}{L^\frac{1}{3} 
	\Gamma \left(-\frac{1}{3}\right)}+\frac{i^{n+m} \sin
	   \left(\frac{1}{4} (4 L-2 \pi  n+2 \pi  m+\pi )\right)}{\sqrt{2 \pi } \sqrt{L}}, 
\end{equation}
where $\Gamma(x)$ is the Euler Gamma function. 
It is important to observe that the first term in~\eqref{eq:c-fin} exhibits 
the decay as $1/L^{1/3}$, whereas the second one is $\propto 1/L^{1/2}$. The first 
term originates from the fact that for $l=\pm1$ the saddle point contribution in the 
integral in~\eqref{eq:int} vanishes. 
The second one is the saddle point contribution 
for $l=0$. 
By using~\eqref{eq:c-fin} it is now straightforward to derive the behavior of  
entanglement entropy of a finite subsystem in the limit $L\to\infty$. As for the 
mutual information~\cite{alba.2019}, to understand the behavior of the entanglement 
entropy at the revival time in the large $L$ limit, one can consider a subsystem 
of length $\ell=2$. An explicit calculation shows that $\delta S$ (cf.~\ref{fig0} (a)) 
decays as $1/L^{2/3}$. 
It is important to observe that the exponent $1/3$ is the same arising in the 
description of super-diffusion on the light-cone in free fermion models. This 
is not surprising because the derivation outlined above is the same.

\end{document}